\documentclass[11pt,a4paper]{article}

\usepackage{amsmath,amsfonts,amssymb}
\usepackage{amscd}
\usepackage[dvipdfmx]{graphicx}
\usepackage{color}
\usepackage{braket}
\usepackage{stmaryrd}
\usepackage{mathrsfs}
\usepackage{slashed}

\usepackage{jheppub}	% JHEP

\makeatletter
\renewcommand{\theequation}{\arabic{section}.\arabic{equation}}
\@addtoreset{equation}{section}

\if0

\newlength{\minitwocolumn}
\fi

\def\DGO{\slashed{\mathcal{D}}}

\newfont{\mmfrak}{eufm10 scaled 1200}
\newcommand{\mfrak}[1]{\mbox{\mmfrak #1}}

\def\be{\begin{equation}}
\def\ee{\end{equation}}
\def\bea{\begin{eqnarray}}
\def\eea{\end{eqnarray}}
\def\benu{\begin{enumerate}}
\def\eenu{\end{enumerate}}
\def\half{{1\over 2}}
\def\p{\partial}
\def\braket#1{\langle#1\rangle}

\def\proof{\medskip\noindent{\sl Proof\ } :\  }

\def\Matrix#1#2{\left(\begin{array}{#1}#2\end{array}\right)}

\def\GL{{\mfrak L}}

\def\CR{{\mathcal R}}
\def\CL{{\mathcal L}}
\def\CG{{\mathcal G}}
\def\CJ{{\mathcal J}}

\def\CH{{\mathcal H}}
\def\CZ{{\mathcal Z}}
 
\def\Sa{{\mathsf a}}
\def\Sb{{\mathsf b}}
\def\Sc{{\mathsf c}}
\def\Sd{{\mathsf d}}

\def\SS{{\mathsf S}}

\def\WM{{\mathbb M}}
\begin{document}
\begin{titlepage}
\begin{flushright}
\null \hfill Preprint TU-1161\\[3em]
\end{flushright}

\begin{center}
{\Large \bf
Metric Algebroid and Poisson-Lie T-duality in DFT
}
\if0
\author{Ursula Carow-Watamura,}
\author{Kohei Miura and}
\author{Satoshi Watamura}

\affiliation{Particle Theory and Cosmology Group,\\Department of Physics, Graduate School of Science, Tohoku University,\\Aoba-ku, Sendai 980-8578, Japan}

\emailAdd{ursula@tuhep.phys.tohoku.ac.jp}
\emailAdd{miura@tuhep.phys.tohoku.ac.jp}
\emailAdd{watamura@tuhep.phys.tohoku.ac.jp}

\fi

\vskip 1.2cm

Ursula Carow-Watamura\footnote{E-mail:~ursula@tuhep.phys.tohoku.ac.jp}, 
Kohei Miura\footnote{E-mail:~miura@tuhep.phys.tohoku.ac.jp}, 
 Satoshi Watamura\footnote{E-mail:~watamura@tuhep.phys.tohoku.ac.jp},

\vskip 0.4cm
{
\it
Particle Theory and Cosmology Group, \\
Department of Physics, Graduate School of Science, \\
Tohoku University \\
Aoba-ku, Sendai 980-8578, Japan \\ %[5ex]

\vskip 0.4cm
}

\vskip 0.4cm

\end{center}

\begin{abstract}
~In this article we investigate the gauge invariance and duality properties of DFT based on a metric algebroid formulation given previously in \cite{2020CWMWY}. The derivation of the general action given in this paper does not employ the section condition. 
Instead, the action is determined by requiring a pre-Bianchi identity on the structure functions of the metric algebroid and also for the dilaton flux.  The pre-Bianchi identity is also a sufficient condition for a generalized Lichnerowicz formula to hold. 
The reduction to the $D$-dimensional space is achieved by a dimensional reduction of the fluctuations. 
The result contains the theory on the group manifold, or the theory extending to the GSE, depending on the chosen background.  
As an explicit example we apply our formulation to the Poisson-Lie T-duality in the effective theory on a group manifold. It is formulated as a $2D$-dimensional diffeomorphism including the fluctuations.
\end{abstract}

%\arxivnumber{22XX.XXXX}

\medskip


\end{titlepage}

\newpage

\tableofcontents

\def\doubleM{{\mathbb M}}
%\begin{document}
\if0
\begin{flushright}
\null \hfill Preprint TU-1101\\[3em]
\end{flushright}
\fi
%\maketitle
\flushbottom
\setcounter{tocdepth}{3}
\section{Introduction}

In the effort to find a geometrical understanding of supergravity and string theory, algebroid structures have received increasing attention in the recent decades. Supergravity can be formulated in various ways, depending on how one treats the gauge symmetry. In generalized geometry \cite{Hitchin:2005in,Gualtieri:aa} the diffeomorphisms and the gauge symmetry of the $B$-field are treated in a geometrical way as generalized diffeomorphisms. Since the $T$-duality in string theory interchanges the metric and the $B$-field, this unification is very natural. 
The algebraic structure of the generalized diffeomorphisms is known to be a Courant algebroid \cite{Courant_1990}.
In the generalized geometry formulation of supergravity  
\cite{Coimbra_2011,Coimbra:2012aa},
the parameter of an infinitesimal transformation of a generalized diffeomorphism is a generalized vector, and the commutator of two infinitesimal transformations defines the Courant bracket. 
%\mc{This is in analogy with general relativity, where the parameter of an infinitesimal transformation of a diffeomorphism is a vector, and the commutator of two infinitesimal transformations defines the Lie bracket 
%of the vector field.}
For reviews see \cite{Hitchin_2011,Bouwknegt_2010}.

Double Field Theory (DFT) \cite{Siegel_1993,Hull:2009aa,Hohm_2010} has been proposed as a T-duality invariant formulation of supergravity.
There, the theory is formulated on a doubled spacetime. It is favourable from the T-duality point of view
 to consider a space, the Fourier modes of which correspond to the winding sector of string theory.  
See \cite{Zwiebach_2012,Aldazabal_2013} for review and further references on DFT.

In the literature DFT is usually formulated on flat space with an $O(D,D)$ metric and a section condition, i.e. a condition on the space of the field configuration. The theory is constructed with the requirement that it becomes a gauge theory after the section condition is applied. 
Here, we denote this formulation of Double Field Theory by DFT$_{sec}$. 
The section condition has often been criticized since it is not
formulated at a Lagrangian level, such as a constraint imposed by a Lagrange multiplier
or a constraint by a gauge symmetry. 
First efforts to address the role of the section condition have been given in \cite{Dibitetto_2012,Aldazabal_2013,Geissbuhler:2013uka}.

From the string theory point of view the section condition is easy to accommodate conceptually, i.e. it can be considered as an equivalent to the matching condition. In DFT, on the other hand, it is the condition to obtain the field theory of the half-dimensional spacetime. However, the section condition does not completely exclude the dual coordinate dependence. This causes an ambiguity in the counting of the degrees of freedom of the resulting theory.  
In this paper, we extend the framework of the DFT using the metric algebroid structure \cite{Vaisman_2012}.
And for this purpose we are forced to be strict with the configuration space of the resulting field theory. Therefore, we apply the dimensional reduction on the field to obtain the field theory of the half-dimensional spacetime. We will see that with this method it becomes possible to analyze the mapping of the fields induced by the
Poisson-Lie T-duality in the effective field theory.

In the formulation of an algebroid, the use of a graded symplectic manifold, the 
so-called QP-manifold 
\cite{Roytenberg:1999} is very convenient, (see also \cite{Cattaneo_2001,Ikeda_2017} and reference therein),  
and its application to DFT has been discussed in 
\cite{Deser_2015,Carow_Watamura_2016,Deser_2018}.
Later, it was shown that a pre-QP manifold is very useful to clarify the algebraic structure of the generalized diffeomorphisms in DFT  \cite{Carow_Watamura_2019}. 

In the QP-manifold approach, the dilaton is not included a priori due to the grading.  
To include it, we use a Clifford bundle where the grading is $Z_2$ 
which allows us to include the dilaton into the Dirac generating operator (DGO) 
\cite{AlekseevDGO}.
The DGO plays the same role as the homological function in the QP-manifold, i.e.,  
it defines the algebroid structure using the derived bracket.
With the use of the DGO the dilaton can be included naturally as shown in \cite{2020CWMWY}, 
due to an ambiguity in the DGO which is not related to the algebroid structure of the generalized vectors.  From the point of view of physics we need a spin structure for supergravity to formulate the fermionic sector and the R-R sector, therefore, the use of the Dirac operator is natural.

As mentioned above, as an algebraic structure of "gauge transformation", we employ the metric algebroid 
and define a generalized DFT action without referring to the section condition. 
We achieve this by means of a generalized Bianchi identity, called pre-Bianchi identity, which is a condition on the structure functions of the metric algebroid \cite{2020CWMWY}.

In this paper we continue our analysis with the aim to formulate the Poisson-Lie T-duality in this framework.
We use the property that the metric algebroid can be reduced to a Courant algebroid by a dimensional reduction of the fluctuations in metric and dilaton.
Technically, this is achieved  
by taking the metric algebroid such that the structure functions 
coincide with the structure functions of a Lie algebroid in a certain frame. Together with the pre-Bianchi identity,
the bracket in the metric algebroid reduces to the Courant bracket after the dimensional reduction of
the fluctuations.
Since our action is constructed in terms of the structure functions of the metric algebroid, after dimensional reduction the action is given in terms of the structure functions of the Courant algebroid. This Courant algebroid is the algebraic structure of the 
generalized diffeomorphisms in the sense of generalized geometry. Thus, it is guaranteed that 
the resulting theory is gauge invariant under the generalized diffeomorphisms on the
reduced spacetime. 

In our approach, we show that the generalized supergravity equation (GSE) \cite{Tseytlin_2016,ARUTYUNOV2016262} can be derived
by taking appropriate structure functions of the metric algebroid.

In the literature \cite{10.1093/ptep/ptx067,Sakatani_2017}, the GSE is derived from DFT$_{sec}$ by adding a generalized vector to the dilaton flux.
When the generalized vector is constant, it is possible to include a linear dual-coordinate dependence in the dilaton while keeping the section condition into the DFT$_{sec}$. 
To include a generalized vector which is not constant, a modification of DFT$_{sec}$, denoted here as DFT${}^{mod}_{sec}$, has been proposed where  
a generalized vector is added directly to the dilaton flux. 
However, the justification of this modification, i.e. the consistency with the algebroid structure, is not addressed.

As we will show, in the formulation given in this paper the form of the dilaton flux is defined geometrically and the modified DFT$_{sec}^{mod}$ is included as a special case of the background, thus filling in the missing algebraic background for the modified DFT action.

To construct an action scalar from the Dirac operator, we have previously established a generalized Lichnerowicz formula 
which follows when the pre-Bianchi identity is satisfied \cite{2020CWMWY}.  
Originally, the Lichnerowicz formula is based on the fact that the square 
of the Dirac operator contains the scalar curvature, which is the Einstein-Hilbert action of gravity.  
Here, we show the generalized Lichnerowicz formula using the DGO and formulate a general action scalar for DFT from the metric algebroid which possesses one free parameter. 
The supergravity action, the DFT action including GSE and the DFT${}_{WZW}$ action can be derived by particular choices of this parameter.  

In the second part of this paper, as an explicit application we investigate the Poisson-Lie T-duality on a Drinfel'd Double as a $2D$ dimensional diffeomorphism.

The organization of this paper is the following: 

In section 2, we summarize definitions of the algebroid structures involved and introduce notations.
We also introduce a new viewpoint concerning the algebraic structure for DFT and collect some results obtained in \cite{2020CWMWY} relevant for the present paper. 

In section 3, we recall the formulation of the Dirac generating operator of the metric algebroid. 

In section 4, the generalized Lichnerowicz formula is given. Then, its decomposition into the subspaces of positive/negative modes is formulated, and details on the dilaton and integration measure are explained.  

While in our previous paper we concentrated on the derivation of the gauge invariant DFT action, here 
we give a more general approach in section 5, starting from the gauge transformations of the corresponding fields. Based on this, the formulation of a general gauge invariant action for DFT is derived. 
In the derivation of this general gauge invariant action no reference to the section condition is made.

Then, we give the relation of this general DFT action and the covariant curvature tensor of the MA, Ricci tensor and scalar curvature which hold as usual.
The action for Supergravity, the DFT action including the GSE and the action for the DFT Wess-Zumino-Witten model are obtained as particular choices of a free parameter in the general DFT action. 
In the last part of this section we show how the dimensional reduction of the fluctuations in metric and dilaton works out. 

In section 6, after a brief summary of the structure of a Drinfel'd Double and the Poisson-Lie T-duality for the case of the sigma model, we give a derivation of Poisson-Lie T-duality for the case of DFT on a Drinfel'd double. In order to establish this duality, the fluctuations on the full doubled space have to be considered in order to establish Poisson-Lie T-duality. This result shows another merit of the metric algebroid formulation.

In section 7, we give the construction of the R-R sector for the DFT action, show its reduction to the DFT$_{sec}$, as well as its properties under local Lorentz transformation and Poisson-Lie T-duality. 

Discussion and conclusions are given in section 8.

\section{Algebroid structures}

We briefly recall here the definitions and introduce notations of Lie algebroid,  Courant algebroid \cite{Courant_1990} and metric algebroid \cite{Vaisman_2012} for convenience. Then, we give an 
improved description of the algebroid structure for DFT. For further aspects of the algebroid structure in DFT see also \cite{Chatzistavrakidis_2018,Mori:2020ut} and references therein. 

\subsection{Lie algebroid}
A Lie algebroid consists of a vector bundle $E \rightarrow M$ over a base manifold $M$, an anti-symmetric bracket $[\cdot,\cdot]_L:E\times E\rightarrow E$ and a bundle map (anchor) $\rho:E\rightarrow TM$ to the tangent bundle $TM$, 
satisfying the following relations 
for $\forall a,b,c\in E$:
\bea
{}[a,fb]_L&=&\rho(a)(f)b+f[a,b]_L~,\\
{}[a,[b,c]_L]_L&=&[[a,b]_L,c]_L+[b,[a,c]_L]_L~.
\eea
The anchor maps the algebroid bracket to the standard Lie bracket $[-,-]_{TM}$ on the tangent bundle $TM$ as
\be
\rho([a,b]_L)= [\rho(a),\rho(b)]_{TM}.
\ee

\subsection{Courant algebroid} 
A Courant algebroid (CA) is a kind of double of a Lie algebroid \cite{Liu_1997}, given by a vector bundle $E\rightarrow M$, endowed with  
a bracket $[-,-] : E\times E\rightarrow E$, 
 a bundle map (anchor) $\rho$ to the tangent bundle $TM$, 
$\rho: E\rightarrow TM$, and a non-degenerate symmetric fiber metric
$\braket{-,-} $ satisfying the three conditions: 
\begin{alignat}{3}
&~~~~~&\rho(a)\braket{b,c}&=\braket{[a,b],c}
+\braket{b,[a,c]}~,
\\&~~~~~&[a,a]&=\half\p\braket{a,a}~,
\\ &~~~~~&[a,[b,c]]&=[[a,b],c]+[b,[a,c]]~, \label{LeibnizidentityofCA}
\end{alignat}
for $a,b,c\in E$ and $f\in C^\infty(M)$,
where a differential $\p$ is defined as $\braket{\p f,a}=\rho(a)f$.

The bracket is not necessarily anti-symmetric up to a derivative term, and the Jacobi identity in Leibniz form of the bracket holds. 
Some useful properties of the bracket can be derived from the above defining equations: 
\begin{alignat}{5}
&[a,fb]&=&~(\rho(a)f)b+f[a,b], \label{property d of CA}~,
\\
&[fa,b]&=&~-(\rho(b)f)a+(\p f)\braket{a,b}+f[a,b]~,\label{property e of CA}\\
&[\p f,a]&=&~0~,\label{property f of CA}\\
&\rho(\p f)&=&~0~,\label{property g of CA}\\
&\rho([a,b])&=&~[\rho(a),\rho(b)]_{TM}~,\label{pCompatibilityOfanchor1}
\end{alignat}
where the bracket $[-,-]_{TM}$ is the standard Lie bracket on $TM$. The bracket on $E=TM\oplus T^*M$ is called the Dorfman bracket.

\subsection{Metric algebroid}
A metric algebroid (MA) $(E,[-,-],\braket{-,-},\rho)$ \cite{Vaisman_2005} is a wider structure than a CA with the Leibniz rule relaxed.
It consists of a vector bundle $E\rightarrow{M}$ endowed with a bracket $[-,-]: E \times E\rightarrow E$,
an inner product $\braket{-,-}: E \times E \rightarrow C^\infty({M})$,
a bundle map (anchor) $\rho:E\rightarrow T{M}$, and a differential $\p$ such that $\braket{\p f,a}=\rho(a)f$ satisfying
\begin{alignat}{3}
&(a)~~~~~&\rho(a)\braket{b,c}&=\braket{[a,b],c}
+\braket{b,[a,c]}~~,
\\&(b)~~~~~&[a,a]&=\half \p\braket{a,a}~~,\label{pCAaxiom2}
\end{alignat}
with $a,b,c\in\Gamma(E)$. Note the absence of the Jacobi identity of the bracket, i.e., the compatibility of the anchor with the bracket (\ref{pCompatibilityOfanchor1}) does not hold in general.

\subsection{Algebroid of DFT revisited}
We construct DFT as a geometry on the vector bundle $E$.  We consider a vector bundle $E$ over a doubled space $\WM$ endowed with a metric algebroid structure as well as a Lie algebroid structure. The maps and brackets are the following:
\bea
\mbox{vector bundle}~\pi&:&E\rightarrow \WM~,\cr
\mbox{LA-bracket(Lie algebroid)}~[\cdot,\cdot]_{L}&:&E\times E\rightarrow E~,\cr
\mbox{MA-bracket(Metric algebroid)}~[\cdot,\cdot]&:&E\times E\rightarrow E~,\cr
\mbox{inner product}~\braket{\cdot,\cdot}&:&E\times E\rightarrow C^\infty(\WM)~,\cr
\mbox{anchor map}~\rho&:&E\rightarrow T \WM~.
\eea
The anchor map $\rho$ is common for both, Lie algebroid and metric algebroid.
We denote fiber metric by $\eta_{AB}$, and the structure functions $F_{ABC}, F'_{ABC}$ and $\phi'_{ABC}$ in a basis $E_A\in E$ are given by
\bea
\eta_{AB}&:=&\braket{E_A,E_B}~,\\
F_{ABC}&:=&\braket{[E_A,E_B],E_C}~,\\
F'_{ABC}&:=&\braket{[E_A,E_B]_L,E_C}~,\\
\phi'_{ABC}&:=&F_{ABC}-F'_{ABC}\label{defphi'}~.
\eea
For DFT we require the $O(D,D)$ structure and thus the fiber metric is a constant $O(D,D)$ metric. 
For later purpose we defined $\phi'$ which measures the difference of the MA bracket and the LA bracket.
The Jacobi identity in the LA gives a relation on the structure functions (Bianchi identity) as follows
\be
\CJ_{ABC}{}^D=\rho(E_{[A})(F'_{BC]}{}^D)+F'_{[AB}{}^EF'_{C]E}{}^D=0~.
\ee
Taking the trace w.r.t. the indices $A$ and $D$, we obtain the relation
\be
\CJ_{ABC}{}^A=2\rho(E_{[B})(\phi'_{A|C]}{}^A)+(\rho(E_A)-\phi'_{DA}{}^D)F_{BC}{}^A-(\rho(E_{A})-\phi'_{DA}{}^D)\phi'_{BC}{}^A=0~.\label{LAJacobi}
\ee
This condition will play a role when we introduce the dilaton in the formulation.

\subsubsection{Jacobiator in Metric Algebroid and Jacobi identity}
In the MA we define a quantity which traces the augmented properties compared to the CA, by defining the following maps $\GL: E\times E\times E\rightarrow E$ 
and $\GL': E\times E \rightarrow E$:
\bea
\GL(a,b,c)&=&[a,[b,c]]-[[a,b],c]-[b,[a,c]] ~,  \label{pCAJacobiator}
\\ \GL'(a,b)&=&[a,b]-[a,b]_L ~,\label{pCAJacobiator2}
\eea
The map $\GL$ in (\ref{pCAJacobiator}) is a Jacobiator in Leibniz-like form.
Note that $\rho\circ\GL'$ does not vanish in general, while in the Courant algebroid, both quantities are zero. 
These quantities satisfy the following relations \cite{2020CWMWY}:
\bea 
\GL(a,b,c)+ \GL(b,a,c)&=&-[\p\braket{a,b},c]~, \label{pSymmetricpartofGLinpreCA}
\\ \GL'(a,b)+\GL'(b,a)&=&\p\braket{a,b}~. \label{psymmetricpartofGL'}
\eea
The above maps are not $C^\infty(M)$-linear in all arguments, however, we can capture their tensorial properties by introducing a map $\phi$
\be
\phi(a,b,c,d)=\braket{\GL(a,b,c),d}~.\label{mapGLtoCM}
\ee
the tensorial property of which is given by
\bea
\Delta \phi
&:= &\phi(fa,b,c,d)+\phi(a,gb,c,d)+\phi(a,b,hc,d)+\phi(a,b,c,kd)-(f+g+h+k)\phi(a,b,c,d)
\cr&=&-(\rho\circ\GL'(b,c)f)\braket{a,d}
+(\rho\circ\GL'(a,c)g)\braket{b,d}
-(\rho\circ\GL'(a,b)h)\braket{c,d}
\cr&&-\braket{a,b}  (\rho\circ\GL'(c,d)f)
+\braket{a,c}(\rho\circ\GL'(b,d)f)
-\braket{b,c}(\rho\circ\GL'(a,d)g)~.\label{ptensorialpropertyofphi}
\eea

Note that the map $\GL'$ given in our previous paper \cite{2020CWMWY} corresponds to the map $\rho\circ\GL'$ in the present paper, since
\be
\rho\circ\GL'(a,b)=\rho([a,b]-[a,b]_L)=\rho([a,b])-[\rho(a),\rho(b)]_{TM}.
\ee
Therefor, in the CA the map $\GL=0$ from which $\rho\circ\GL'f=0$ follows. 
%From the %since $[\p f,a]=0$ in (\ref{p_usefulrelationinpreCA}). 
The MA reduces to a CA if $\GL=0$. Now, the pre-Courant algebroid \cite{Vaisman_2005, BRUCE2019254} is characterized by the condition on the MA $\GL\not=0$, but $\rho\circ\GL'f=0$. 

Note that from eq. (\ref{defphi'}) it follows that 
\be
{\cal L}'(E_A,E_B) = \phi'_{AB}{}^CE_C~.\label{phi'def}
\ee

\section{Dirac generating operator}

A Dirac generating operator is a Dirac operator which generates the algebroid structure. 
Similarly as done in the generalized geometry \cite{AlekseevDGO,Garcia-Fernandez:2016aa,Severa_2017} we can formulate the MA structure by using a Dirac generating operator. For this end, we consider a Clifford bundle $Cl(E)$ endowed with a graded bracket $\{\cdot,\cdot\}$. 
The generators $\gamma_A$ of the Clifford bundle satisfy 
\be
\{\gamma_A,\gamma_B\}=2\eta_{AB}~,
\ee
where $\eta_{AB}$ is the $O(D,D)$ metric.
The basis of $Cl(E)$ is given by the products
\be
\{1,\gamma_{A_1},\gamma_{A_1A_2},\cdots,\gamma_{A_1A_2\cdots A_{2D}}\}\in Cl(E)~,
\ee
where $\gamma_{A_1A_2\cdots A_n}$ is
\be
\gamma_{A_1A_2\cdots A_n}=\gamma_{[A_1}\gamma_{A_2}\cdots \gamma_{A_n]}~.
\ee
$\gamma_A$ is related to $E_A$ by a linear map $\gamma:E\rightarrow Cl(E)$,
\be
\gamma(E_A)=\gamma_A~,
\ee
and for $^\forall a,b\in E$,
\be
\{\gamma(a),\gamma(b)\}=2\braket{a,b}~.
\ee
In the following we do not explicitly write the map $\gamma$ to avoid cumbersome notation, i.e.,
$\gamma_A=E_A$ and $E\subset Cl(E)$.
A representation space of $Cl(E)$ is given by a $O(D,D)$ spin bundle $\mathbb{S}$.
For the spin bundle we introduce a basis $\gamma_A=(\gamma^a,\gamma_a)$ which satisfies
\be
\{\gamma^a,\gamma_b\}=2\delta^a_b~.
\ee
The vacuum state of $\mathbb{S}$ is defined in the standard way, 
\be
\gamma_a\Ket{0}=0~,
\ee
and a ${2D}$-dimensional spinor is defined by 
multiplying $\gamma^a$ onto the vacuum.
We define a derivative $\partial_A$ acting on $\mathbb{S}$ by 
\bea
\{\partial_A,\gamma_B\}&=&0~,\\
\{\partial_A,f\}&=&\rho(E_A)(f)~,\\
\partial_A\Ket{0}&=&0~.
\eea
The operations of the metric algebroid are generated by the Dirac generating operator via the derived bracket \cite{Kosmann-Schwarzbach:2004aa} as
\bea
\{\{\DGO,a\},b\}&=&[a,b]~,\\
\{\DGO,f\}&=&\frac{1}{2}\partial f~,\\
\{\{\DGO,a\},f\}&=&\rho(a)(f)~. 
\eea
A concrete form of $\DGO$ for the MA has been given in \cite{2020CWMWY} as
\be
\DGO=\frac{1}{2}\gamma^A\partial_A-\frac{1}{24}\gamma^{ABC}F_{ABC}-\frac{1}{4}\gamma^AF_A~,
\ee
where $\gamma^{ABC} = \gamma^{[A}\gamma^B\gamma^{C]}$ and $F_A$ is an ambiguity of $\DGO$ which is not determined by the axioms.
A priory, the flux $F_A$ has nothing to do with a dilaton in this level. To obtain a condition which connects $F_A$ to the dilaton we need to require further structure, e.g., scale invariance or introduction of a line bundle or a determinant bundle \cite{AlekseevDGO,Severa:2018aa}. 

\paragraph{Compatible connection}
We can represent the Dirac generating operator $\DGO$ using a covariant derivative on $\mathbb{S}$.
The covariant derivative on $\mathbb{S}$ is given by a spin connection $W_{ABC}$,
\be
\nabla_A^{\mathbb{S}}=\partial_A-\frac{1}{4}\gamma^{BC}W_{ABC}~.
\ee
We can choose the spin connection such that
\be
\DGO=\frac{1}{2}\gamma^A\nabla_A^\mathbb{S}~,
\ee
becomes the Dirac generating operator by the following condition 
\bea
T_{ABC}&:=&3W_{[ABC]}-F_{ABC}=0~,\\
W^B{}_{BA}{}&=&F_A~,
\eea
where $T_{ABC}$ is a generalized torsion, i.e., we obtain a torsion-free connection \cite{Garcia_Fernandez_2014}.

\section{Generalized Lichnerowicz formula and generalized dilaton}

\subsection{Generalized Lichnerowicz formula}
In a previous paper \cite{2020CWMWY} we have shown that in the MA formulation DFT belongs to a class which is specified by a generalized Bianchi identity, which we called pre-Bianchi identity, for the structure functions $F_{ABC}$.  We also found that the pre-Bianchi identity for $F_{ABC}$ and for $F_A$ are the necessary and sufficient conditions for a generalized Lichnerowicz formula to hold.
\footnote{For the original Lichnerowicz formula see \cite{Lichnerowicz1963,Bismut_1989}.  An application to M-theory has been formulated in \cite{Coimbra_2019a}.
}
We defined a generalized Lichnerowicz formula for the MA by using the Dirac generating operator.
%Here, we show that the generalized Lichnerowicz formula gives a condition on the flux $F_A$. 

In generalized geometry, an action has been constructed by the square of the Dirac operator 
$\DGO^2$ \cite{Coimbra_2011,Coimbra:2012aa},\cite{Severa:2018aa},\cite{Coimbra_2019a}.

In DFT on the other hand, $\DGO^2$ contains second order and first order derivative terms.
In order to derive a scalar from $\DGO^2$, we subtract 
a particular Laplacian using $\phi'$ as follows.

We observe that $\phi'_{ABC}$ changes under the local $O(D,D)$ transformation 
in the same way as the spin connection $W_{CAB}$, i.e., under the local $O(D,D)$ transformation of the basis
\be
\delta_\Lambda E_A=\Lambda_A{}^B E_B~,
\ee
they transform as
\bea
\delta_\Lambda W_{CAB}&=&\rho(E_C)(\Lambda_{AB})+\Lambda\triangleright W_{CAB}~,\\
\delta_\Lambda \phi'_{ABC}&=&\rho(E_C)(\Lambda_{AB})+\Lambda\triangleright\phi'_{ABC}~,\label{phi'transformation}
\eea
where $\Lambda\triangleright$ means the linear term of the transformation, e.g., 
\be
\Lambda\triangleright\phi'_{ABC}=\Lambda_A{}^{A'}\phi'_{A'BC}+\Lambda_B{}^{B'}\phi'_{AB'C}+\Lambda_C{}^{C'}\phi'_{ABC'}~.
\ee
\proof Recall that in the present formulation $\phi'_{ABC}$ is introduced as the difference of MA and LA bracket. 
The tensorial property of $\phi'_{ABC}$ is then given by
\if0
\bea
\Delta\braket{[a,b],c}
&:=&\braket{[fa,b],c}+\braket{[a,gb],c}+\braket{[a,b],hc}-
(f+g+h)\braket{[a,b],c}
\cr &=&\braket{a,b}\rho(c)f-\braket{a,c}\rho(b)f
+\braket{b,c}\rho(a)g~.
\eea
\bea
\Delta\braket{[a,b]_L,c}
&:=&\braket{[fa,b]_L,c}+\braket{[a,gb]_L,c}+\braket{[a,b]_L,hc}
\cr&&~~~~~~~~~~~~~~~~~~~~~~-(f+g+h)\braket{[a,b]_L,c}
\cr &=&-\braket{a,c}\rho(b)f
+\braket{b,c}\rho(a)g~.
\eea
\fi
\bea
\Delta\braket{[a,b]-[a,b]_L,c}&:=&\braket{[fa,b]-[fa,b]_L,c}+\braket{[a,gb]-[a,gb]_L,c}+\braket{[a,b]-[a,b]_L,hc}
\cr&&~~~~~~~~~~~~~~~~~~~~~~~~~~~~~~~~-(f+g+h)\braket{[a,b]-[a,b]_L,c}
\cr&=&\braket{a,b}\rho(c)f
\eea
Thus, in the basis $E_A$ this gives 
\be
\Delta\braket{[E_A,E_B]-[E_A,E_B]_L,E_C}=\eta_{AB}\rho(E_C)f
\ee
and we obtain (\ref{phi'transformation}).
\if0
\be
\delta_\Lambda \phi'_{ABC}=\delta_\Lambda \braket{[E_A,E_B]-[E_A,E_B]_L,E_C}=\rho(E_C)\Lambda_{AB}+\Lambda\triangleright
\phi'_{ABC}
\ee
\fi

We use this fact to define a second covariant derivative $\nabla_A^{\phi'}$ as
\be
\nabla_A^{\phi'}=\partial_A-\frac{1}{4}\gamma^{BC}\phi'_{BCA}~,
\ee
from which we construct a Laplacian $\Delta^{\phi'}$
\be
\Delta^{\phi'}=\eta^{AC}(\nabla_A^{\phi'}-\phi'_{BA}{}^B-U_A)\nabla^{\phi'}_C=div_{\nabla^{\phi'}}^U\nabla^{\phi'}~.
\ee
The quantity $U=U^A E_A$ is an arbitrary generalized vector $U$ appearing as an ambiguity in the divergence $div_{\nabla^{\phi'}}^U$. See also \cite{Severa:2018aa}.

Then, the generalized Lichnerowicz formula can be obtained by requiring that
\be
4\DGO^2-\Delta^{\phi'}\in C^\infty(M)~,\label{conditionL}
\ee
i.e., this difference should yield a function.
In the general case, 
\bea
4\DGO^2-\Delta^{\phi'}&=&
-{1\over24}F_{ABC}F^{ABC}-\half(\rho(E^A)F_A)+(-F^A+\phi'_{E}{}^{AE}+U^A)\p_A
+{1\over4}F_AF^A
+{1\over8}\phi'_{BCA}\phi'^{BCA}
\cr&&
+{1\over 4}\left(-\CJ_{BCD}{}^D
+ 2(\rho(E_{[B})(-F_{C]}+\phi'^D{}_{C]D})-(-F^A+\phi'_D{}^{AD})F_{ABC}-U_A\phi'_{BC}{}^A
\right)\gamma^{BC}
\cr&&-{1\over48}\Big(4\rho(E_{[A})(F_{BCD]})-3F_{[AB}{}^EF_{CD]E}+3\phi'_{[AB}{}^E\phi'_{CD]E}\Big)\gamma^{BCB'C'}
%\cr=&
%{1\over4}R^\nabla_{AB}{}^{AB}
%+(-W_B{}^{BA}+\phi'_{B}{}^{AB}+U^A)\p_A
%%+{1\over8}\phi'_{BCA}\phi'^{BCA}
%\cr&
%-{1\over 4}\Big(
%\CJ_{BCD}{}^D
%+ (\rho(E_{[B})(W^B{}_{B|C]}-\phi'^D{}_{C]D})-(W_B{}^{BA}-\phi'_D{}^{AD})F_{ABC}+U_A\phi'_{BC}{}^A
%\Big)\gamma^{BC}
%\cr&
%-{1\over48}\tilde\phi_{BCB'C'}\gamma^{BCB'C'}~,
\label{pGeneralizedLichnerowicz1}
\eea
where $\CJ_{BCD}{}^D$ is the tensor defined in (\ref{LAJacobi}).

From the requirement (\ref{conditionL}) i.e., the difference gives the scalar, the terms proportional to 
$\partial_A,\gamma^{AB}$ and $\gamma^{ABCD}$ must vanish. This requirement gives a condition for $F_A$ and $U_A$:
\be
(F_A-\phi'_{BA}{}^B-U_A)\rho(E^A)=0~,\label{fluxA}
\ee
Together with  the condition from the term proportional to $\gamma^{AB}$, 
\be
\rho(U_AE^A)=2\rho(E_A)(d)~,
\ee
This defines $F_A$ with an ambiguity $U'_A$
\be
F_A=2\rho(E_A)d+\phi'_{BA}{}^B  +U'_A
\ee
where 
\bea
\rho(U'_AE^A)&=&0~,\\
2\rho(E_{[B})(U'_{C]})&=&F_{BC}{}^AU'_A~,
\eea
and $d\in C^\infty(M)$ is an arbitrary scalar, which
can be interpreted as a generalized dilaton if we introduce further structure. 
When $E_A{}^M$ is invertible, $U'_A=0$.\footnote{In DFT$_{sec}$, the vector bundle is identified with the tangent bundle itself, thus $E_A{}^M$ is invertible.}
If $E\neq TM$, $U'_A$ is not automatically zero. 
Here, we choose that the degree of freedom of $F_A$ is given by the scalar $d$ only, i.e., 
\bea
F_A&=&\phi'_{BA}{}^B+2\rho(E_A)(d)~,\\
U'_A&=&0~.
\eea
Using this solution of $F_A$ we obtain
\bea
4\DGO^2-\Delta^{\phi'}
&=&
-\frac{1}{24}F_{ABC}F^{ABC}-\frac{1}{2}\rho(E^A)(F_A)
+\frac{1}{4}F_AF^A
+\frac{1}{8}\phi'_{BCA}\phi'^{BCA}
\cr
&&-\frac{1}{4}\Big(2\rho(E_{[B})(F_{C]})+\rho(E_A)(F_{BC}{}^A)-F_AF_{BC}{}^A+\rho(E_A)(\phi'_{BC}{}^A)-F_A\phi'_{BC}{}^A
\Big)\gamma^{BC}
\cr
&&-{1\over48}\Big(4\rho(E_{[A})(F_{BCD]})-3F_{[AB}{}^EF_{CD]E}+3\phi'_{[AB}{}^E\phi'_{CD]E}\Big)\gamma^{ABCD}~.
\eea
The term proportional to $\gamma^{AB}$ is the pre-Bianchi identity including the flux $F_A$,
\be
\mathcal{B}_{BC}:=2\rho(E_{[B})(F_{C]})+\rho(E_A)(F_{BC}{}^A)-F_AF_{BC}{}^A+\rho(E_A)(\phi'_{BC}{}^A)-F_A\phi'_{BC}{}^A=0~,
\label{pre-Bianchi_with_dilaton}
\ee
and the term proportional to $\gamma^{ABCD}$ is the pre-Bianchi identity
\be
\mathcal{B}_{ABCD}:=4\rho(E_{[A})(F_{BCD]})-3F_{[AB}{}^EF_{CD]E}+3\phi'_{[AB}{}^E\phi'_{CD]E}=0~.
\label{pre-Bianchi_without_dilaton}
\ee
for the flux $F_{ABC}$ (which corresponds to the tensor $\tilde\phi$ in \cite{2020CWMWY}).

Therefore, the requirement that (\ref{conditionL}) is a scalar gives the 3 conditions (\ref{fluxA}), (\ref{pre-Bianchi_with_dilaton}) and (\ref{pre-Bianchi_without_dilaton}), and 
the scalar part of the generalized Lichnerowicz formula yields
\be
4\DGO^2-\Delta^{\phi'}=-\frac{1}{24}F_{ABC}F^{ABC}-\frac{1}{2}\rho(E^A)(F_A)
+\frac{1}{4}F_AF^A
+\frac{1}{8}\phi'_{BCA}\phi'^{BCA}~. \label{generalLich}
\ee
This is the desired scalar which is $O(D,D)$ invariant. However, it is not the action of DFT 
since we have not yet used the data associated with the generalized metric $H_{AB}$.
To obtain the action for DFT, we have to construct an $O(1,D-1)\times O(D-1,1)$ invariant scalar.

\subsection{Projected Lichnerowicz formula}

After the introduction of the generalized metric $H_{AB}$ on the fiber, the local $O(D,D)$ symmetry reduces to 
$O(1,D-1)\times O(D-1,1)$ in such a way that the invariance of the action under generalized diffeomorphism is preserved at least up to a closure constraint.
The projected Lichnerowicz formula gives  $O(1,D-1)\otimes O(D-1,1)$ invariant Lagrangians.

The generalized metric on the fiber $H_{AB}$ is 
an $O(D,D)$ symmetric tensor and satisfies
\be
H_{AB}\eta^{BC}H_{CD}=\eta_{AD}.\label{ODDpropertyofmetric}
\ee
As in the generalized geomety, the generalized metric gives the split of the tangent bundle to the positive/negative subbundle as $E = V^+ \oplus V^-$:
\bea
V^+ &=& \{ V\in T\mathbb M | H^{A}{}_B V^B = V^A\}~, \cr
 V^- &=& \{ V\in T\mathbb M | H^{ A}{}_B V^B = -V^{B}\}
\eea
where $H^A{}_B=\eta^{AC}H_{CB}$. Since $H^A{}_B H^B{}_C =\delta^A{}_B$,
we have corresponding projectors given by 
\be
P^{\pm A}{}_B = {1\over 2}(\delta^A{}_B \pm H^A{}_B)~,
\ee
which satisfy
\be
P^{+A}{}_B + P^{-A}{}_B =\delta^A{}_C ~, ~ P^{\pm A}{}_B  P^{\pm B}{}_C = P^{\pm A}{}_C~ ,~P^{\pm A}{}_B
P^{\mp B}{}_C =0~.
\ee

Correspondingly, we consider the projected spin bundle  
$\Gamma(\mathbb{S}^\pm)$ which is a module over $Cl(V^\pm)$, constructed on $\Ket{0}$ by multiplying the elements of $\Gamma(V^\pm)$\cite{2020CWMWY}.
The Dirac generating operator is also split by using the covariant derivatives
 $\nabla^{{\mathbb S}^\pm}:\Gamma({\mathbb S}^\pm) \rightarrow \Gamma(E^*)\otimes\Gamma({\mathbb S}^\pm)$ and denoted as $\DGO^\pm$.

The projected Lagrangians $\mathscr L^\pm$ are given by 
\be
\mathscr L^{\pm} = 4(\DGO^{\pm})^{2} + div_{\nabla}\nabla_{\mp}^{S^\pm} -\Delta^{\phi' \pm}\ , \label{projectLich}
\ee
where $\Delta^{\phi' \pm}$ is the divergence of the projected covariant derivative $\nabla^{{\phi'}^\pm}$. For details see \cite{2020CWMWY}.
The explicit form of ${\cal L}^\pm$ is given as:
\be
\mathscr L^+ = -\frac{1}{24}F_{\Sa\Sb\Sc}F^{\Sa\Sb\Sc}-\frac{1}{8}F^{\bar{\Sa}\Sb\Sc}F_{\bar{\Sa}\Sb\Sc}-\frac{1}{2}\rho(E_{\Sa})F^{\Sa}+\frac{1}{4}F_{\Sa} F^{\Sa}+\frac{1}{8}\phi'_{\Sa\Sb\Sc} \phi'^{\Sa\Sb\Sc} + \frac{1}{8}\phi'_{\Sa\Sb{\bar \Sc}}\phi'^{\Sa\Sb {\bar \Sc}}
\ee
and
\be
\mathscr L^-=-\frac{1}{24}F_{\bar{\Sa}\bar{\Sb}\bar{\Sc}}F^{\bar{\Sa}\bar{\Sb}\bar{\Sc}}-\frac{1}{8}F^{\Sa{\bar \Sb}{\bar \Sc}}F_{\Sa{\bar \Sb}{\bar \Sc}}-\frac{1}{2}\rho(E_{\bar{\Sa}})F^{\bar{\Sa}}+\frac{1}{4}F_{\bar{\Sa}}F^{\bar{\Sa}}+\frac{1}{8}\phi'_{\bar{\Sa}\bar{\Sb}\bar{\Sc}}\phi'^{\bar{\Sa}\bar{\Sb}\bar{\Sc}} + \frac{1}{8}\phi'_{\bar{\Sa}\bar{\Sb}{\Sc}}\phi'^{\bar{\Sa}\bar{\Sb}{\Sc}}
\ee
Here, the bases of the positive/negative subspaces are denoted 
as $(E_\Sa, E_{\bar \Sa})$ 
i.e., the indices of the corresponding fluxes are $\Sa /\bar \Sa$, respectively.

\subsection{Dilaton and Measure}

The generalized Lie derivative ${\cal L}_a$ is the generator of the generalized diffeomorphisms in $M$.
On the spin bundle $\mathbb{S}$ using a Dirac generating operator 
$\DGO$ the Lie derivative can be formulated 
for all $a\in E$ and all $\chi\in \mathbb{S}$ as
\be
{\cal L}_a\chi =\{\DGO,a\}\chi~,
\ee
which is compatible with the generalized Lie derivative on $E$, i.e., the Leibniz rule holds for all 
$a, b \in E,\chi\in \mathbb{S}$
\be
{\cal L}_ab\chi=\{\DGO,a\}b\chi=\{\{\DGO,a\}b\}\chi+b\{\DGO,a\}\chi=({\cal L}_ab)\chi+b{\cal L}_a\chi~.
\ee
This gives a gauge transformation of a spinor with weight $1\over 2$ \cite{2020CWMWY}.
We define an inner product $(\chi_1,\chi_2)_A$ on $\mathbb{S}$ which is invariant under the infinitesimal $O(D,D)$ transformation,  and satisfies  
for all $\chi_1,\chi_2\in \Gamma(\mathbb{S})$:

\be
(\gamma_A\chi_1,\chi_2)_A=(\chi_1,\gamma_A\chi_2)_A%\int d\mu 
%\chi_1^\dagger A_+\chi_2~,
\ee
\be
(\DGO\chi_1,\chi_2)_A=(\chi_1,-\DGO\chi_2)_A~.
\ee
In this paper, we call $(\cdot,\cdot)_A$ simply an $A$-product. (See the appendix for details).
The $A$-product is an inner product such that the anti-hermiticity of the Dirac operator 
holds.

Then, we can compute
\be
(\{\DGO,\gamma_A\}f_1\Ket{0},f_2 K\Ket{0})_A=(f_1\Ket{0},-\{\DGO,\gamma_A\}f_2 K\Ket{0})_A~,
\ee
for all $f_1,f_2\in C^\infty(M)$. The $K$ in front of a vacuum is introduced in (\ref{def_hatK}) to define the dual vacuum as (\ref{dualvacuumofspinor}). 
By a straightforward calculation, we get 
\be
((\partial_A-\frac{1}{2}F_A)f_1\Ket{0},f_2 K\Ket{0})_A=(f_1\Ket{0},-(\partial_A-\frac{1}{2}F_A)f_2 K\Ket{0})_A~.
\label{partintdilatonmeasure}
\ee
This is a necessary condition when we define the integration measure of the $A$-product.

A concrete form of the $A$-product can be derived by assigning for $f\in C(\WM)$ and vacuum
 $(\Ket{0}, K\Ket{0})_A$
\bea
(\Ket{0}, f K\Ket{0})_A&=&\int d\mu f~,\\
d\mu &=& dX {\tilde{h}}~.
\eea
Then, the measure $\tilde{h}$ has to satisfy
\bea
{\tilde{h}}^{-1}\rho(E_B)({\tilde{h}})&=&-\partial_NE_B{}^N-\phi'_{AB}{}^A-2\rho(E_B)d\cr
&=&-\partial_NE_B{}^N+F'_{AB}{}^A-2\rho(E_B)d~,
\eea
where $\rho(E_A)=E_A{}^N\partial_N$. Thus, we factor out the function $d$
 as ${\tilde{h}}=e^{-2d}h$. 
To obtain an explicit form for $h$ 
we require that the trace of the flux on the LA satisfies
\be
F'_{AB}{}^C(\delta_C{}^A-E_C{}^N\eta_{NM}E{}^{AM})=0~,
\ee
where $\eta_{MN}$ is an induced metric on $TM$,
\bea
\eta^{MN}&:=&E_A{}^M\eta^{AB}E_B{}^N~,\\
\eta_{LM}\eta^{MN}&=&\delta_L{}^N~.
\eea
Then, we can calculate $h$ as follows
\bea
h^{-1}\rho(E_B)(h)&=&-\partial_NE_B{}^N+F'_{AB}{}^A\cr
&=&-\partial_NE_B{}^N+F'_{AB}{}^CE_C{}^N\eta_{NM}E_D{}^M\eta^{DA}\cr
&=&-\partial_NE_B{}^N+(E_A{}^L\partial_LE_B{}^N-E_B{}^L\partial_LE_A{}^N)\eta_{NM}E_D{}^M\eta^{DA}\cr
&=&-\partial_NE_B{}^N+\partial_NE_B{}^N-E_B{}^L\partial_LE_A{}^N\eta_{NM}E_D{}^M\eta^{DA}\cr
&=&-\frac{1}{2}E_B{}^L\partial_LE_A{}^N\eta_{NM}E_D{}^M\eta^{DA}-\frac{1}{2}E_B{}^L\partial_LE_D{}^N\eta_{NM}E_A{}^M\eta^{DA}\cr
&=&\frac{1}{2}E_A{}^N\rho(E_B)(\eta_{NM})E_D{}^M\eta^{DA}\cr
&=&\frac{1}{2}\eta^{NM}\rho(E_B)(\eta_{NM})\cr
&=&\sqrt{\det\eta_{NM}}^{-1}\rho(E_B)(\sqrt{\det\eta_{NM}})~.
\eea
This yields a concrete form for the measure in $(\Ket{0},K\Ket{0})_A$ as
\bea
h&=&c_0\sqrt{\det\eta_{NM}}~,\\
{\tilde{h}}&=&c_0e^{-2d}\sqrt{\det\eta_{NM}}~,\\
(\Ket{0},K\Ket{0})_A&=&c_0\int dXe^{-2d}\sqrt{\det\eta_{NM}}~.
\eea
where $c_0$ is a constant.

The transformation of the scalar $d$ under the generalized Lie derivative is obtained as follows. 
We consider the $A$-product of the vacuum $\Ket{0}$ and its dual $K\Ket{0}$, 
%i.e., K\gamma_AK^{-1}=H_A{}^B\gamma_B$.
\bea
{\delta}_a(\Ket{0},fK \Ket{0})_A&=&({\cal L}_a\Ket{0},fK\Ket{0})_A+(\Ket{0},f{\cal L}_aK\Ket{0})_A\cr
&=&(\{\DGO,a\}\Ket{0},fK\Ket{0})_A+(\Ket{0},f\{\DGO,a\}K\Ket{0})_A\cr
&=&(\Ket{0},K\Ket{0})_Af(\rho(E_A)-F_A)(a^A)~,\\
{\delta}_a\int dX{\tilde{h}}f&=&\int dXc_0 {\delta}_ae^{-2d}\sqrt{\det\eta_{MN}}f\cr
&=&\int dXc_0 (-2{\delta}_ad)e^{-2d}\sqrt{\det\eta_{MN}}f\cr
&=&(\Ket{0},K\Ket{0})_A(-2{\delta}_ad)f~.
\eea
Therefore, the transformation rule of the scalar $d$ is determined as
\bea
{\delta}_a d&=&-\frac{1}{2}(\rho(E_A)-F_A)(a^A)~,\label{dilatontrans}\\
{\delta}_a e^{-2d}&=&(\rho(E_A)-F_A)(a^A)e^{-2d}~.
\eea

The last equation reflects the gauge transformation of the dilaton. 
Therefore, we identify $d$ with the dilaton.

\section{DFT actions and gauge symmetry}

First, we define a NS-NS field $U_A{}^B$ as the fluctuation of the generalized vielbein.
We assume that all physical degrees of freedom of the generalized vielbein
are equal to those of the generalized metric $H_{MN}=E_M{}^AE_N{}^BH_{AB}$.
This means that $\eta_{MN}=E_M{}^AE_N{}^B\eta_{AB}$ is not physical.

For a given $\eta_{MN}$, we define a background vielbein $\bar E_A{}^M$ such that 
 the $O(D,D)$ metric in the background basis $\bar E_A{}^N\partial_N=\bar E_A$ 
is given by
\be
\bar\eta_{AB}:=\bar E_A{}^M\bar E_B{}^N\eta_{MN}=
\begin{pmatrix}
&\delta^a{}_b\\
\delta_a{}^b&
\end{pmatrix}~.
\ee
Then, we can separate the generalized vielbein $E_A{}^M$ into a fluctuation $U_A{}^B$ and
the background $\bar E_B{}^N$ as
\be
E_A{}^N=U_A{}^B\bar E_B{}^N~,\label{splitfluctuation}
\ee
As we will show, $\bar E_A{}^N$ and $U_A{}^B$ relate to $\eta_{MN}$ and $H_{MN}$ respectively.
This $O(D,D)$ metric $\bar \eta_{AB}$ and the $O(D,D)$ metric $\eta_{AB}$ 
in the local Lorentz basis are connected by $U_A{}^B$ as
\be
U_A{}^{A'}U_B{}^{B'}\bar\eta_{A'B'}=\eta_{AB}~.
\ee
Since $\eta_{AB}=\bar\eta_{AB}$, $U_A{}^B\in O(D,D)$.
Moreover, considering that the degrees of freedom of $U_A{}^B$ which do not change $H_{MN}$ are not physical, the space of fluctuations $U_A{}^B$ is given by
\be
U_A{}^B\in (O(1,D-1)\times O(D-1,1))\backslash O(D,D)~.
\ee
The concrete form of $U_A{}^B$ can be written by
\be
E_A=U_A{}^B\bar E_B~,~U_A{}^B=
\begin{pmatrix}
e^{-T}{}^a{}_b&0\\
-e_a{}^cB_{cb}&e_a{}^b
\end{pmatrix}~,
\ee
where $e_a{}^b$ has $O(1,D-1)$ symmetry and $B_{ab}$ is an anti-symmetric tensor.
In $DFT_{sec}$, 
%$e_a{}^b$ and $B_{ab}$ relate to the $D$-dimensional vielbein $e_a{}^m$and the B field $B_{mn}$ respectively, and 
the background $\bar E_A{}^M$ is flat, i.e.,
$\bar E_A{}^N$ can be chosen as $\delta_A{}^N$.

\subsection{Action in MA formalism}
The most general $O(1,D-1)\otimes O(D-1,1)$ invariant action is given by a linear combination 
of $\mathscr L^+$ and $\mathscr L^-$ as
\bea
S=\cal I(\beta_+,\beta_-) &=&\beta_+\Big(-4(\DGO^+\Ket{0},\DGO^+K\Ket{0})_A 
- (\nabla^{S^+}_{E_{\bar{a}}}\Ket{0},\nabla^{S^+}_{E^{\bar{a}}}K\Ket{0})_A 
+(\nabla^{(\phi')^+}_{E_{A}}\Ket{0},\nabla^{(\phi')^+}_{E^{A}}K\Ket{0})_A\Big)
\cr&+&\beta_-\Big(-4(\DGO^-\Ket{0},\DGO^-K\Ket{0})_A 
- (\nabla^{S^-}_{E_a}\Ket{0},\nabla^{S^-}_{E^a}K\Ket{0})_A 
+(\nabla^{(\phi')^-}_{E_{A}}\Ket{0},\nabla^{(\phi')^-}_{E^{A}}K\Ket{0})_A\Big)~,
\cr&=& (\Ket{0},K\Ket{0})_A (\beta_+ {\mathscr L}^+ +\beta_-{\mathscr L}^-)
\cr&=& c_0 \int dX \sqrt{\det\,\eta }e^{-2d} (\beta_+\mathscr L^+ +\beta_-\mathscr L^-)~.\label{Beta_action}
\eea
The coefficients $\beta_+$ and $\beta_-$ are free parameters. 
\begin{itemize}
\item Our requirement is that after the dimensional reduction we obtain the supergravity action.  
This requirement leads to a condition on the parameters $\beta_\pm$ as
\be
-\beta_++\beta_-=8c_0^{-1}~,
\ee
 where the overall constant $c_0$ is fixed such that we obtain the standard normalization of the Einstein-Hilbert action.

\item To discuss the Poisson-Lie T-duality in DFT, we choose the action with the following parameters 
\be
S_{DFT}={\cal I}(0,8c_0^{-1})\ .  \label{action_PLTd}
\ee
As we shall see, the GSE is naturally included in this parametrization.

\item From the action (\ref{action_PLTd}) the DFT$_{sec}$ action can also be produced by requiring that the MA-bracket on the coordinate basis vanishes, 
\be
[\partial_L,\partial_M]=0~.
\ee

\item The action of the DFT Wess-Zumino-Witten model $S_{DFT_{WZW}}$ discussed in \cite{Blumenhagen:2014gva,Blumenhagen_2015,BosqueHasslerL2016} can be constructed when the MA-bracket equals to the
LA-bracket in the basis $\bar E_A$:
\bea
[\bar E_A,\bar E_B]&=&[\bar E_A,\bar E_B]_L~.\\
\bar F_{\Sa\Sb\Sc}&=&f_{\Sa\Sb}{}^\Sd s_{\Sd\Sc}~,~\bar F^{\bar\Sa\bar\Sb\bar\Sc}=-\bar{f}^{\bar\Sa\bar\Sb}{}_{\bar\Sd}s^{\bar\Sd\bar\Sc}~,~{\rm others}=0~,
\eea
The action of the DFT$_{WZW}$ model is given by
\be
S_{DFT_{WZW}}={\cal I}(-4c_0^{-1},4c_0^{-1})~.
\ee
\end{itemize}

\if0
Up to the overall coefficients this action coincides with the one derived in our previous paper [ ]
\bea
S&=&\int d^{2D}x\sqrt{\det \eta}e^{-2d}\Big(-\frac{1}{24}F_{abc}F^{abc}-\frac{1}{8}F^{\bar{a}bc}F_{\bar{a}bc}-\frac{1}{2}\rho(E_a)F^a+\frac{1}{4}F_aF^a+\frac{1}{8}\phi'_{abC}\phi'^{abC}\Big)
\cr&=&\int d^{2D}xf\Big(-\frac{1}{24}F_{abc}F^{abc}-\frac{1}{8}F^{\bar{a}bc}F_{\bar{a}bc}-\frac{1}{2}\rho(E_a)F^a+\frac{1}{4}F_aF^a+\frac{1}{8}\phi'_{abC}\phi'^{abC}\Big)
\eea

With this form of the action \rc{(5.16)} several models in physics can be discussed depending on the choice of the coefficients $\beta_\pm$.
\fi

\paragraph{Riemann tensor and DFT action}

In a previous paper we formulated the generalized curvature ${\cal R}(a,b,c,d)$ on the
metric algebroid.  The result is 
\be
{\cal R}(a,b,c,d) = {\cal R}^\nabla(a,b,c,d)+{\cal R}^\nabla(c,d,a,b) + \braket{\GL'(a,b),\GL'(c,d) }%_{TM}
\ee
where
\be
{\cal R}^\nabla(a,b,c,d):=
\braket{(\nabla^E_{a}\nabla^E_{b}-\nabla^E_{b}\nabla^E_{a})c-\nabla^E_{[a,b]}c,d}
+\half\braket{\nabla^E_{E_A}a,b}\braket{\nabla^E_{E^A}c,d}~.
\ee
In the basis $E_A$ the generalized Riemann tensor is given as 
\bea
{\cal R}_{ABCD} &:= &{\cal R}(E_A,E_B,E_C,E_D)
\cr&=& {\cal R}^\nabla_{ABCD} + {\cal R}^\nabla_{CDAB} + \phi'_{ABE}\phi'_{CD}{}^E ~,
\label{generalizedRiemanntensor}
\eea
where 
\be
{\cal R}^\nabla_{ABCD}:= 2\rho (E_{[A})W_{B]CD} - 2W_{[A|C}{}^EW_{|B]ED} 
-F_{AB}{}^EW_{ECD}+{1\over 2}W_{EAB}W^E{}_{CD} ~.
\ee

From the generalized Riemann tensor ${\cal R}_{ABCD}$, we can construct the various 
generalized Ricci tensors and curvature scalar by using the projection operators
made by the $O(D,D)$ metric $\eta_{AB}$ and the $O(1,D-1)\times O(D-1,1)$ invariant generalized metric $H_{AB}$. The above action can be expressed by using the projected generalized 
Riemann scalar as 
\bea
\mathscr L^+&=&\frac{1}{8}\CR_{\Sa\Sb}{}^{\Sa\Sb},\\
\mathscr L^-&=&\frac{1}{8}\CR_{\bar\Sa\bar\Sb}{}^{\bar\Sa\bar\Sb}.
\eea
Thus the Lagrangian of the most general action is
\be
\CL=\beta_+\mathscr L^+ +\beta_-\mathscr L^- =\frac{1}{8}\beta_+\CR_{\Sa\Sb}{}^{\Sa\Sb}+\frac{1}{8}\beta_-\CR_{\bar\Sa\bar\Sb}{}^{\bar\Sa\bar\Sb}
\ee

\subsection{Variation of the action}
\paragraph{Variation with respect to the vielbein}
In this subsection, we consider a variation of the DFT action (\ref{Beta_action})
by an infinitesimal $O(D,D)$ rotation. 
The variation of the vielbein $\delta^{(E)}$ is defined by
\be
\delta^{(E)} E_A=\Lambda_{A}{}^B E_B
\ee
where $\Lambda_{AB}=-\Lambda_{BA}$. 
The variation of the fluxes yields
\begin{align}
\delta^{(E)} F_{ABC}&=\rho(E_A)(\Lambda_{BC})+\rho(E_B)(\Lambda_{CA})+\rho(E_C)(\Lambda_{AB})+3\Lambda_{[A}{}^{A'}F_{A'|BC]}\\
\delta^{(E)} \phi'_{ABC}&=\rho(E_C)(\Lambda_{AB})+\Lambda_A{}^{A'}\phi'_{A'BC}+\Lambda_B{}^{B'}\phi'_{AB'C}+\Lambda_C{}^{C'}\phi'_{ABC'}~,\\
\delta^{(E)} F_A&=\rho(E_B)(\Lambda^B{}_A)+\Lambda_A{}^BF_B~.
\end{align}
The measure is invariant
\be
\delta^{(E)} (dX\sqrt{\det\eta}e^{-2d})=0~.
\ee
Since the action is given by the projection corresponding to positive/negative generalized tangent vectors, we also split the infinitesimal transformation parameter accordingly as
\be
\Lambda_{AB}\rightarrow \{\Lambda_{\Sa\Sb},\Lambda_{\Sa\bar \Sb},\Lambda_{\bar \Sa\bar \Sb}\}\ .
\ee

Then, the variation of the $\mathscr L^+$ part of the action is
\newcommand{\cod}[2]{\overset{\mbox{\tiny $#1$}}{#2}{}}
\newcommand{\codt}[4]{\overset{\mbox{\tiny $#1\!\!#2\!\!#3$}}{#4}{}}
\begin{align}
&\delta^{(E)} \Big(c_0\int dX\sqrt{\det\eta}e^{-2d}{\mathscr L}^+\Big)\cr
&=c_0\int dX\sqrt{\det\eta}e^{-2d}\cr
&~~~\Big[-\frac{1}{4}{\Lambda}^{\Sa\Sb}\Big(({E}_C^N\partial_N-{F}_C){F}_{\Sa\Sb}{}^C
+2{E}_{[\Sa}^N\partial_N{F}_{\Sb]}-({E}_C^N\partial_N-{F}_C){\phi'}_{\Sa\Sb}{}^C\Big)\cr
&~~~+\frac{1}{2}{\Lambda}^{\Sa {\bar \Sd}}\Big({F}_{\Sa}{}^{\Sb{\bar \Sc}}{F}_{{\bar \Sd} \Sb{\bar \Sc}}-({E}_\Sc^N\partial_N-{F}_\Sc){F}_{\Sa\bar\Sd}{}^{\Sc}+{E}_{\bar \Sd}^N\partial_N{F}_\Sa-{\phi'}_\Sa{}^{\Sb\Sc}{\phi'}{\bar\Sd \Sb\Sc}-{\phi'}_\Sa{}^{\Sb\bar\Sc}{\phi'}_{\bar\Sd\Sb\bar\Sc}\Big)\Big]\cr
%&=c_0\int dX\sqrt{\det\eta}e^{-2d}\Big[-\frac{1}{4}{\Lambda}^{ab}{\mathcal{B}}_{ab}\cr
%&~~~+\frac{1}{2}{\Lambda}^{a\bar d}\Big({F}_a{}^{b\bar c}{F}_{\bar d b\bar c}-({E}_c^N\partial_N-{F}_c){F}%_{a\bar d}{}^c+{E}_{\bar d}^N\partial_N{F}_a-{\phi'}_a{}^{bc}{\phi'}_{\bar d bc}-{\phi'}_a{}^{b\bar c}{\phi'}_{\bar d %b\bar c}\Big)\Big]\cr
&=c_0\int dX\sqrt{\det\eta}e^{-2d}\cr
&~~~\frac{1}{2}{\Lambda}^{\Sa\bar \Sd}\Big({F}_\Sa{}^{\Sb\bar \Sc}{F}_{\bar\Sd \Sb\bar \Sc}-({E}_\Sc^N\partial_N-{F}_\Sc){F}_{\Sa\bar \Sd}{}^\Sc+{E}_{\bar \Sd}^N\partial_N{F}_\Sa-{\phi'}_\Sa{}^{\Sb\Sc}{\phi'}_{\bar\Sd \Sb\Sc}-{\phi'}_\Sa{}^{\Sb\bar\Sc}{\phi'}_{\bar\Sd \Sb\bar\Sc}\Big)~.\label{var_Lplus}
\end{align}
Here we have used that the term proportional to $\lambda^{\Sa\Sb}$ is exactly the pre-Bianchi identity ${\mathcal{B}}_{AB}$ in eq. (\ref{pre-Bianchi_with_dilaton}) and thus vanishes. Therefore, the nontrivial variation is given by the terms proportional to $\lambda^{\Sa\bar\Sb}$ only, which gives the above result. 
 
For ${\mathscr L}^-$ we compute correspondingly:
\begin{align}
&\delta^{(E)} \Big(c_0\int dX\sqrt{\det\eta}e^{-2d}{\mathscr L}^-\Big)\cr
%&=c_0\int dX\sqrt{\det\eta}e^{-2d}\cr
%&~~~\frac{1}{2}{\Lambda}^{\bar a d}\Big({F}_{\bar a}{}^{\bar bc}{F}_{d\bar b c}-({E}_{\bar c}^N\partial_N-{F}_{\bar c}){F}_{\bar ad}{}^{\bar c}+{E}_{d}^N\partial_N{F}_{\bar a}-{\phi'}_{\bar a}{}^{\bar b\bar c}{\phi'}_{d\bar b\bar c}-{\phi'}_{\bar a}{}^{\bar b c}{\phi'}_{d\bar b c}\Big)~.\cr
&=-c_0\int dX\sqrt{\det\eta}e^{-2d}\cr
&~~~\frac{1}{2}{\Lambda}^{\Sa\bar\Sd}\Big({F}_{\bar\Sd}{}^{\bar\Sb\Sc}{F}_{\Sa\bar\Sb \Sc}-({E}_{\bar \Sc}^N\partial_N-{F}_{\bar \Sc}){F}_{\bar\Sd \Sa}{}^{\bar \Sc}+{E}_{\Sa}^N\partial_N{F}_{\bar \Sd}-{\phi'}_{\bar \Sd}{}^{\bar\Sb\bar\Sc}{\phi'}_{\Sa\bar\Sb\bar \Sc}-{\phi'}_{\bar \Sd}{}^{\bar\Sb \Sc}{\phi'}_{\Sa\bar\Sb \Sc}\Big)~.\label{var_Lminus}
\end{align}
Again, the term proportional to $\Lambda_{\bar\Sa\bar\Sb}$ vanishes due to the pre-Bianchi identity and therefore, the nontrivial part of the variation is proportional to $\Lambda_{\Sa\bar\Sb}$ only.
We shall see that the above variation by the $O(D,D)$ transformation gives the equation of motion of the generalized vielbein for DFT on a Drinfeld double.

\paragraph{Variation with respect to the dilaton}
We also need the variation of the generalized dilaton $d\rightarrow d+\delta d$, which gives for the fluxes 
\bea
\delta^{(d)}F_{ABC} &=& 0~,\cr
\delta^{(d)}\phi'_{ABC} &=& 0~,\cr
\delta^{(d)}F_A &=& 2\rho (E_A)(\delta d)
\eea
and for the measure
\be
\delta^{(d)}(dX\sqrt{\det \eta} e^{-2d})(-2\delta d) = \sqrt{\det \eta} e^{-2d}(-2\delta d)~.
\ee
Thus, the variation of the action with respect to the generalized dilaton is
\be
\delta^{(d)}S =c_0 \int dX \sqrt{\det \eta}\ e^{-2d}\delta d(-2\beta_+\ {\mathscr L}^+ -2\beta_-\ {\mathscr L}^-)~.
\ee

\if0
which can be expressed by using eq. \ref{curvature_scalar} as
For $\beta_+=-\beta_-$, $\beta_+-\beta_-=-8c_0^{-1}$
\be
\delta^{(d)}S_{DFT} = \int dX \sqrt{\det \eta}\ e^{-2d}(\delta d) {\cal R}~.
\ee
from this follows eom (420) for dilaton
\be
{\cal R} = 0~.
\ee
\fi

\subsection{Variation and Ricci tensors}

The terms proportional to $\Lambda^{\Sa\bar\Sd}$ in eqs. (\ref{var_Lplus}) and (\ref{var_Lminus}) are related to the Ricci tensor also in the metric algebroid approach.
From the generalized Riemann tensor (\ref{generalizedRiemanntensor}),
we can formulate two types of generalized Ricci tensor by taking the trace with the projection,
namely: 
\bea
\CR_{\Sa\bar\Sa}^+&=&2\CR_{\Sa\Sb\bar\Sa}{}^{\Sb},\\
\CR_{\bar\Sa\Sa}^-&=&2\CR_{\bar\Sa\bar\Sb\Sa}{}^{\bar\Sb}
\eea
Explicitly, they are 
\bea
\CR_{\Sa\bar\Sa}^+&=&\CR_{\bar\Sa\Sa}^+=-2(F_{\Sa}{}^{\Sb\bar\Sc}F_{\bar\Sa \Sb\bar\Sc}-(\partial_{\Sc}-F_{\Sc})F_{\Sa\bar\Sa}{}^{\Sc}+\partial_{\bar\Sa}F_{\Sa}-\phi'_{\Sa}{}^{\Sb C}\phi'_{\bar\Sa\Sb C})\\
\CR_{\bar\Sa\Sa}^-&=&\CR_{\Sa\bar\Sa}^-=-2(F_{\bar\Sa}{}^{\bar\Sb\Sc}F_{\Sa \bar\Sb\Sc}-(\partial_{\bar\Sc}-F_{\bar\Sc})F_{\bar\Sa\Sa}{}^{\bar\Sc}+\partial_{\Sa}F_{\bar\Sa}-\phi'_{\bar\Sa}{}^{\bar\Sb C}\phi'_{\Sa\bar\Sb C})
\eea
Comparing with (\ref{var_Lplus}) and (\ref{var_Lminus}), we can write
\bea
\delta^{(E)}\int dX h\CL^+&=&\frac{1}{4}\int dX h\Lambda^{\Sa\bar\Sa}\CR_{\Sa\bar\Sa}^+\\
\delta^{(E)}\int dX h\CL^-&=&-\frac{1}{4}\int dX h\Lambda^{\Sa\bar\Sa}\CR_{\Sa\bar\Sa}^-
\eea
Thus, for the most general action we get
\bea
\delta^{(E)}{\cal I}(\beta_+,\beta_-)=\delta^{(E)}\int dX h(\beta_+\CL^++\beta_-\CL^-)&=&2\int dX h\Lambda^{\Sa\bar\Sa}(\frac{1}{8}\beta_+\CR_{\Sa\bar\Sa}^+-\frac{1}{8}\beta_-\CR_{\Sa\bar\Sa}^-)
\eea

\paragraph{Equations of motion}

The above variation gives also the equation of motion for vielbein and dilaton.

\subsection{Local gauge symmetry}

As in generalized geometry, the gauge transformation of vielbein 
is generated by the generalized Lie derivative with weight. Here, we denote the gauge parameter which is a generalized vector by $V\in T\WM$
\be
\delta_V E_A ={\cal L}_V E_A=[V,E_A]=\Lambda_A{}^B E_B~, 
\ee
where
\bea
\Lambda_{AB} &=& \rho(E_B) V_A - \rho(E_A) V_B - V^C F_{CBA}~.
%\cr\delta d &\rightarrow& -{1\over 2} (\partial_A -F_A) V^A~.
\eea
Therefore, we can use the result for the general variation.
The gauge transformation of the generalized dilaton $d$ is given in (\ref{dilatontrans}).

Thus, we obtain the gauge variation of the action as 
\bea
\delta_V S_{DFT} &=&c_0 \int dX \sqrt{\det \eta} e^{-2d}\cr
&&{\beta_+\over 8}\Big ( -2
(\partial_\Sa V_{\bar \Sb} - \partial_{\bar \Sb}V_\Sa -V^\Sc F_{\Sc\Sa\bar \Sb} - V^{\bar \Sc}F_{\bar \Sc \Sa\bar \Sb})
{\cal R}^{+{\Sa\bar \Sb}}+ (\partial_A -F_A)V^A{\cal R}_{\Sa\Sb}{}^{\Sa\Sb} \Big )
\cr&&+{\beta_-\over 8}\Big ( 2
(\partial_\Sa V_{\bar \Sb} - \partial_{\bar \Sb}V_\Sa -V^\Sc F_{\Sc\Sa\bar \Sb} - V^{\bar \Sc}F_{\bar \Sc \Sa\bar \Sb})
{\cal R}^{-{\Sa\bar \Sb}}+ (\partial_A -F_A)V^A{\cal R}_{\bar\Sa\bar\Sb}{}^{\bar\Sa\bar\Sb} \Big )
\eea

\if0

****************

\bea
\delta_V S_{DFT} &=& \int dX \sqrt{\det \eta} e^{-2d}\cr
&&\Big ( 2
(\partial_\Sa V_{\bar \Sb} - \partial_{\bar \Sb}V_\Sa -V^\Sc F_{\Sc\Sa\bar \Sb} - V^{\bar \Sc}F_{\bar \Sc \Sa\bar \Sb})
{\cal R}^{-{\Sa\bar \Sb}}+ (\partial_A -F_A)V^A{\cal R}_{\bar\Sa\bar\Sb}{}^{\bar\Sa\bar\Sb} \Big )
%\cr
%&=& \int dX \sqrt{\det \eta} e^{-2d}\cr
%&&{\Big [} -2V^c {\Big (} (\partial_{\bar \Sb}-F_{\bar \Sb}){\cal R}_\Sc{}^{\bar \Sb} - F_{\Sc\Sa\bar \Sb}{\cal R}^{\Sa\bar \Sb} +{1\over 4}\partial_\Sc{\cal R}{\Big )} \cr
%&&+ 2V^{\bar \Sc} {\Big (} (\partial_\Sa-F_\Sa){\cal R}^\Sa{}_{\bar \Sc} + F_{\bar \Sc \Sa\bar \Sb}{\cal R}^{\Sa\bar \Sb} -{1\over 4}\partial_{\bar \Sc}{\cal R}{\Big )} {\Big ]}~.
\eea

**********
\fi

For the gauge invariance, we require the condition
\be
\delta_V S_{DFT} = 0~,
\ee
after the dimensional reduction.
We find that the corresponding sufficient conditions are given by
\bea
\delta_V F_{BCD}&=&\rho(V)F_{BCD}~,\label{conditionZ4}\\
\delta_V F_A&=&\rho(V)F_{A}~,\label{conditionZ2}\\
\phi'_{ABC}\phi'_{DE}{}^C&=&0~.
\label{gauge_cond_phi'}
\eea
On the other hand, the gauge transformations of the structure functions with gauge parameter $V^A$ are
\bea
\delta_{V}F_{BCD}&=&V^A\partial_AF_{BCD}+\CZ_{VBCD}~,\\
\delta_{V}F_B&=&V^A\partial_AF_B+\CZ_{VB}~,
\eea
where $\CZ_{V BCD}$ and $\CZ_{V A}$ are obtained from the commutator of $V^A$ with the square of the Dirac generating operator as
\be
\{4\DGO^2,V\}=-\gamma^A\CZ_{V A}+2\gamma^A[(\partial^BV_A)+V^C\phi'_{CA}{}^B]\partial_B-\frac{1}{6}\CZ_{V BCD}\gamma^{BCD}~,
\ee
and are defined by
\bea
\CZ_{VBCD}&=&\frac{1}{2}\{\{\{\{\DGO^2,V\},E_B\},E_C\},E_D\}~,\\
\CZ_{VB}&=&-(\partial^A-F^A)\partial_AV_B-\phi'_{CB}{}^A\partial_AV^C-V^A(\partial_C-F_C)\phi'{}_{AB}{}^C~.
\eea
Therefore, conditions (\ref{conditionZ4}) and (\ref{conditionZ2}) imply that
\bea
\CZ_{V BCD}&=&0~,\label{gauge_cond_FABC}\\
\CZ_{V A}&=&0~.\label{gauge_cond_FA}
\eea
Thus (\ref{gauge_cond_phi'}), (\ref{gauge_cond_FABC}) and (\ref{gauge_cond_FA}) are the 
conditions for the gauge invariance.
\footnote{Note that using the pre-Bianchi identity, in the local Lorentz basis the above conditions reduce to the Bianchi identities 
 \bea
(\partial_C-F_C)F_{AB}{}^C+2\partial_{[A}F_{B]}&=&(\partial_C-F_C)\phi'_{AB}{}^C=-\CZ_{AB}\\
4\partial_{[A}F_{BCD]}-3F_{[AB}{}^EF_{CD]E}&=&-3\phi'_{[AB}{}^E\phi'_{CD]E}=-\CZ_{ABCD}
\eea
However, since $\CZ_{AB}$ and $\CZ_{ABCD}$ are not tensors, 
this does not mean that the conditions (\ref{gauge_cond_FABC}) and (\ref{gauge_cond_FA}) hold.}

In DFT, we require the gauge invariance only in the $D$-dimensional theory, and 
here we realize it by the dimensional reduction of the fluctuation.
As a special case,
we can construct a gauge invariant DFT which includes DFT$_{sec}$ and DFT$_{WZW}$ as follows.
First, we separate the vielbein into a fluctuation $U_A{}^B\in O(D,D)$ and a background 
$\bar E_A$ as in (\ref{splitfluctuation}).

For a reduction to the $D$-dimensional field theory, we require that the fluctuation $U_A{}^B$ depends only on the $D$-dimensional coordinate $x^m$, where the polarization of the $D$-dimensional coordinate should be 
chosen appropriately for a given $2D$-dimensional manifold. We will show an example for the case of the Drinfel'd double in the following section.
The same reduction condition must be applied for the dilaton $d$ and the gauge parameter $V^A$.

Furthermore,
we have to require that the derivative $\rho(E_A)$ on $U_A{}^B$ and $d$ in the action also depends on $x^m$ only,  
meaning that $\bar E_A{}^m\partial_m$ is required to depend only on $x^m$.

With the above requirement for $U_A{}^B,d$ and $\bar E_A{}^m$, we obtain the action on the
 $D$-dimensional space. To restore the gauge invariance of the D-dimensional action
  originated from the metric algebroid, 
 we consider (\ref{conditionZ4}) and (\ref{conditionZ2}), as well as the condition on the $\phi'_{ABC}$ in (\ref{gauge_cond_phi'}).
 
 We start with the condition (\ref{gauge_cond_phi'})
on  the structure function $\phi'_{ABC}$ which is given by
\be
\phi'_{ABC}=U_A{}^{A'}U_B{}^{B'}U_C{}^{C'}\bar{\phi}'_{A'B'C'}+\rho(E_C)(U_A{}^{A'})U_B{}^{B'}\eta_{A'B'}~.
\ee
The first term can be adjusted by choosing the MA such that in the $\bar E_{A}$ basis
\be
\braket{\GL'(\bar E_A,\bar E_B),\bar E_C}=\bar{\phi}'_{ABC}=0\label{conditionphi'}\ .
\ee
Then, (\ref{gauge_cond_phi'}) holds  
if the $O(D,D)$ metric satisfies 
\be
\eta^{CC'}\rho(E_C)(U_A{}^{A'})\rho(E_{C'})(U_D{}^{D'})=0\ .\label{sectionlike}
\ee
This means that $\eta^{-1}{}^{MN}$ in a local coordinate satisfies
\be
\eta^{-1}{}^{mn}=0~. \label{sectionlikecondition2}
\ee
Note that the condition (\ref{sectionlike}) is not the section condition.  It is a condition 
on the 2D-dimensional manifold. The condition  (\ref{sectionlikecondition2}) depends on
the choice of the local coordinate. We require that 
we can find such a polarization of $x^m$ and a local coordinate system satisfying (\ref{sectionlikecondition2})
by choosing an appropriate coordinate system.
If we can not find such a coordinate system, we can not apply the dimensional reduction 
and thus we exclude such a doubled space from our consideration.

The condition (\ref{sectionlikecondition2}) holds of course in the flat case. We shall see that
in the case of a Drinfel'd double, this condition is also satisfied naturally.

To restore the gauge invariance of the D-dimensional action,
 we must still consider (\ref{conditionZ4}) and (\ref{conditionZ2}).
Since 
the pre-Bianchi identities in the basis $\bar E_A$ reduce to the Bianchi identities
by the condition $\bar{\phi}'_{ABC}=0$,
which we have already required by the choice of the MA, the conditions (\ref{conditionphi'}), (\ref{gauge_cond_FABC}) and (\ref{gauge_cond_FA}) are satisfied after the dimensional reduction. Therefore, (\ref{conditionZ4}) and (\ref{conditionZ2}) also hold.

Since in the usual discussion of the section condition the basis where the closure holds is not clearly specified, its meaning is obscure even in the flat case.
On the other hand, in our approach the condition for the gauge invariance is formulated in the local Lorentz frame, thus it is covariant.
The dimensional reduction is performed with respect to the fluctuation
in a specific local coordinate and therefore, the function space of the dynamical field is well defined.
It means that a dual coordinate dependence is not allowed. This is different from the section condition which allows a dual coordinate dependence in general.

To summarize,  by taking a MA satisfying $\bar\phi'_{ABC}=0$ and requiring that the $2D$-dimensional metric in a local coordinate satisfies $\eta^{-1}{}^{mn}=0$,
the gauge invariance is guaranteed by the above dimensional reduction of the fluctuations.
\footnote{Note that taking the MA such that 
$\bar\phi'_{ABC}=0$ in the $\bar E_A$ frame means
\be
\bar F_{ABC}=\bar{F}'_{ABC}~.
\ee
Since $\bar{F}'_{ABC}$ satisfies the closure condition of the Lie algebroid,
from the pre-Bianchi identity (\ref{pre-Bianchi_without_dilaton}) for $\bar{F}_{ABC}$, we conclude that
\be
\rho(\bar E_{[A})\bar{F}_{BCD]}=0~.
\ee
Thus, we assume that $\bar{F}_{ABC}=\bar{F}'_{ABC}$ is a constant flux in the following.
}

\if0

We take the convention 
\be
\eta_{AB} = \Matrix{cc}{0&1 \cr 1 & 0} \ , \ \ \CH_{AB} = \Matrix{cc}{s^{ab}&0 \cr 0 & s_{ab}}\ ,
\ee 
where $s_{ab}$ is the local Lorentz $O(1,D-1)$ metric.

\fi

\section{Derivation of Poisson-Lie T-duality by DFT action}

Poisson-Lie T-duality was first introduced in \cite{KLIMCIK1995455} as a generalization of 
non-abelian T-duality in the non-linear sigma model. 
In the algebroid context it is discussed in \cite{Severa_2015,JURCO20181,Jurco2018EFTPL,Severa:2018aa}. 
In DFT it has been discussed in  \cite{Blumenhagen:2014gva,Hassler2016TheTO,Hassler:2017yza,Sakatani_2019,Demulder_2019}.
Here, we apply our results of the MA formulation of DFT to the Poisson-Lie T-duality. We briefly recall the Poisson-Lie T-duality for convenience and to introduce notation.

\subsection{Poisson-Lie T-duality}\label{PLTD}
Poisson-Lie T-duality is a generalization of T-duality based on 
a Drinfel'd double. The Poisson-Lie T-duality is formulated in the most clear form by using the $\cal E$ model \cite{KLIMCIK1996116}, which is a special case of a sigma model on a Drinfel'd double.
The Drinfel'd double $\mathcal{D}=\mathcal{G}\bowtie\bar{\mathcal{G}}$ is a $2D$-dimensional group constructed by two $D$-dimensional groups $\mathcal{G},\bar{\mathcal{G}}$.
The corresponding Lie algebras are denoted by $\mathfrak{d}$ and $\mathfrak{g}$, $\bar{\mathfrak{g}}$, respectively. 
Their generators and brackets are denoted as
\be
t_a\in \mathfrak{g}~,~ \bar t^a\in \bar{\mathfrak{g}}~.
\ee
\be
[t_a,t_b]_{\mathfrak{g}}=f_{ab}{}^ct_c~,~[\bar{t}^a,\bar{t}^b]_{\bar{\mathfrak{g}}}=\bar{f}^{ab}{}_c\bar{t}^c~,
\ee
while the generators of $\mathfrak{d}$ are given by
\be
\mathfrak{d} \ni  T_A:=(t_a,\bar t^a) ~.
\ee
The Lie algebra $\mathfrak{d}$ is expressed in terms of the structure constants of $\mathfrak{g}$ and $\bar{\mathfrak{g}}$ as
\be
[t_a,t_b]_{\mathfrak{d}}=f_{ab}{}^ct_c~,
\ee
\be
~[\bar{t}^a,\bar{t}^b]_{\mathfrak{d}}=\bar{f}^{ab}{}_c\bar{t}^c~.
\ee
\be
[t_a,\bar t^b]_{\mathfrak{d}}=\bar f^{bc}{}_at_c-f_{ac}{}^b\bar t^c~.
\ee
The third bracket is due to the $O(D,D)$ structure: 
$\mathfrak{d}$ has 
an inner product defined as
\be
\braket{T_A,T_B}=\eta_{AB}~,
\ee
with the compatibility condition
\be
\braket{T_A,[T_B,T_C]_{\mathfrak{d}}}=\braket{T_B,[T_C,T_A]_{\mathfrak{d}}}~.
\ee
This condition yields the third equation containing both structure constants.

Finally, from the Jacobi identity
\be
[T_A,[T_B,T_C]_{\mathfrak{d}}]_{\mathfrak{d}}+[T_B,[T_C,T_A]_{\mathfrak{d}}]_{\mathfrak{d}}+[T_C,[T_A,T_B]_{\mathfrak{d}}]_{\mathfrak{d}}=0\ ,
\ee
the following condition for the structure constants is obtained
\be
f_{ab}{}^e\bar f^{cd}{}_e=4f_{e[a}{}^{[c}\bar f^{d]e}{}_{b]}~.
\ee

The Poisson-Lie T-duality is 
the equivalence of two non-linear sigma models in different backgrounds specified by the metric $G$ and the $B$-field,
\be
S[X] \sim \int d^2 \sigma [G_{mn}dX^m\wedge *dX^n + B_{mn}dX^m\wedge dX^n]~,
\ee
where the field $X$ is an embedding into the target space.  Formulated in the $\cal E$-model, the field is an element $l\in \mathcal{D}$ of a Drinfel'd double \cite{Klim_k_1996,Klimcik:wy,Klimcik:2015vk}.
Its action is given by
\begin{equation}
S[l]=\frac{1}{2}\int  d^2\sigma\lbrack\braket{l^{-1}\partial_\sigma l,l^{-1}\partial_\tau l}-\braket{l^{-1}\partial_\sigma l,\hat{\mathcal{H}}_0 (l^{-1}\partial_\sigma l)}\rbrack+\frac{1}{12}\int \braket{l^{-1}dl\stackrel{\wedge}{,}[ l^{-1}dl\stackrel{\wedge}{,}l^{-1}dl]}~,
\label{E-model_action}
\end{equation}
where $\hat\CH_0$ is a linear map $\mathfrak{d}\rightarrow\mathfrak{d}$, defined by
\begin{align}
\braket{T_A,\hat{\mathcal{H}}_0(T_B)}=&
	\begin{pmatrix}
		(G_0^{-1})^{ab}&(G_0^{-1}B_0)^a{}_b\\
		-(B_0G_0^{-1})_a{}^b&(G_0-B_0G_0^{-1}B_0)_{ab}
	\end{pmatrix}~.
\end{align}
$G_0$ and $B_0$ in the above equation are constant matrices.
$l$ can be parameterized by $g\in\mathcal{G}$ and $\bar g\in\bar{\mathcal{G}}$ as
\be
l=\bar g(\bar x)g(x)~.
\ee
We denote the local coordinates of $\mathcal{D}=\bar \CG\times \CG$ as 
$X^M=(\bar x_m, x^m)$, respectively.
Since $\bar g$ does not carry physical degrees of freedom in this parameterization, it can  be integrated out. Then, we obtain a non-linear sigma model $\hat S[g]$
with a metric $\hat G$ and a B-field $\hat B$ as:
\begin{equation}
\hat G+\hat{B}=L^{-1}\frac{1}{\frac{1}{G_0+B_0}+\Pi}L^{-T}~,
\end{equation}
where $L^{-1}$ and $\Pi$ are defined by
\begin{align}
	g^{-1}dg=&L^{-1}{}_m{}^at_adX^m~,\label{left_invariant_of_g}\\
	\begin{pmatrix}
		g^{-1}\bar{t}^ag\\
		g^{-1}t_ag
	\end{pmatrix}
	=&
	\begin{pmatrix}
		a^{-T}{}^a{}_b(g)&-a^{-T}{}^a{}_c(g)\Pi^{cb}(g)\\
		0&a{}_a{}^b(g)
	\end{pmatrix}
	\begin{pmatrix}
		\bar{t}^b\\
		t_b
	\end{pmatrix}\nonumber\\
	=&
	\begin{pmatrix}
		a^{-T}{}^a{}_c(g)&0\\
		0&a{}_a{}^c(g)
	\end{pmatrix}
	\begin{pmatrix}
		\delta^c{}_b(g)&-\Pi^{cb}(g)\\
		0&\delta{}_c{}^b(g)
	\end{pmatrix}
	\begin{pmatrix}
		\bar{t}^b\\
		t_b
	\end{pmatrix}~.
\end{align}

On the other hand, using a coordinate transformation $X\rightarrow X'$, the field $l\in\mathcal{D}$ can also be parameterized as
\be
l=g(x')\bar g(\bar x')~.
\ee
with the local coordinates $X'^M=(\bar x'_m, x'^m)$. In this case, $g$ can be integrated out and we obtain an action $\check{S}[\bar g]$ with a metric and a $B$-field given by
\be
\check{G}+\check{B}=\bar L^{-1}\frac{1}{G_0+B_0+\bar \Pi}\bar L^{-T}~,
\ee
where
\begin{align}
	\bar g^{-1}d\bar g=&\bar L^{-1}{}^m{}_a\bar t^ad\bar x^m~,\label{left_invariant_of_gbar}\\
	\begin{pmatrix}
		\bar{g}^{-1}\bar{t}^a\bar{g}\\
		\bar{g}^{-1}t_a\bar{g}
	\end{pmatrix}
	=&
	\begin{pmatrix}
		\bar{a}{}^a{}_b(\bar{g})&0\\
		-\bar{a}^{-T}{}_a{}^c(\bar{g})\bar{\Pi}_{cb}(\bar{g})&\bar{a}^{-T}{}_a{}^b(\bar{g})
	\end{pmatrix}
	\begin{pmatrix}
		\bar{t}^b\\
		t_b
	\end{pmatrix}~. 
	\label{def_bar_a_Pi}
\end{align} 

Since the two actions $\hat S[g(x)]$ and $\check S[\bar g(\bar x')]$ are equivalent at classical level, the non-linear sigma models on the backgrounds
$\hat G+\hat B$ and $\check G+\check B$ are classically equivalent.
This equivalence is called Poisson-Lie T-duality.

\subsection{DFT on Drinfel'd Double}
In this section, we define the DFT on a Drinfel'd double.

We define a vielbein $E_A{}^N$ at a point $X^M$ on $\mathcal{D}$ corresponding to an element $l\in\mathcal{D}$ and a fluctuation $U_A{}^B$ in a local basis
\be
E_A=E_A{}^N(l)\p_N=U_A{}^B(X)\bar E_B{}^N(X)\p_N ~,\label{vielbein_drin}
\ee
where $\bar E_N{}^A(l)$ is defined by the left invariant current of $\mathcal{D}$:
\be
l^{-1}dl=\bar E^{-1}{}_N{}^AT_AdX^N=\bar E^{-1}{}_N{}^at_adX^N+\bar E^{-1}{}_N{}_a\bar t^adX^N~.
\ee
The Lie algebra of the left invariant vector $\bar E_A=\bar E_A{}^N(l)\partial_N$ is given by 
\be
[\bar E_A,\bar E_B]_L=\bar F'_{AB}{}^C\bar E_C~.
\ee
where $\bar F'_{AB}{}^C$ are the structure constants of $\mfrak d$.  Splitting the basis $\bar E_A$ into $\bar E^a$ and $\bar E_a$, we obtain
\bea
{}[\bar E_a,\bar E_b]_L&=&f_{ab}{}^c\bar E_c~,\\
{}[\bar E_a,\bar E^b]_L&=&\bar f^{bc}{}_a\bar E_c-f_{ac}{}^b\bar E^c~,\\
{}[\bar E^a,\bar E^b]_L&=&\bar f^{ab}{}_c\bar E^c~.
\eea

Since in the MA the flux is not determined uniquely 
by the vielbein, we have a freedom to fix
the flux $\bar F_{AB}{}^C$ of the MA in the basis $\bar E_A$ as
\footnote{This means that we are considering a class which allows us to take this choice.
As we shall see this class includes the case on the Drinfeld double.}
\be
[\bar E_A,\bar E_B]=\bar F_{AB}{}^C\bar E_C=\bar F'_{AB}{}^C\bar E_C~.
\ee
Consequently, for this choice the structure function $\bar{\phi}'_{ABC}$ vanishes in the $\bar E_A$ basis
\be
\bar{\phi}'_{ABC} :=\braket{[\bar E_A,\bar E_B]-[\bar E_A,\bar E_B]_L,\bar E_C}=0~. \label{gauge_fixing}
\ee 
The $O(D,D)$ metric is defined by
\be
\bar\eta_{AB}=\braket{\bar E_A,\bar E_B}=
\begin{pmatrix}
0&\delta^a{}_b\\
\delta_a{}^b&0
\end{pmatrix}~.
\ee
With the above $O(D,D)$ metric we have
\be
\braket{\bar E_A,[\bar E_B,\bar E_C]_L}=\braket{\bar E_B,[\bar E_C,\bar E_A]_L}~,
\ee
which is compatible with the MA, i.e., since in the $\bar E_A$ basis $\phi'=0$ the above cyclicity generalizes to the relation on the MA consistently:
\be
-\braket{\bar E_A,[\bar E_C,\bar E_B]}=\braket{\bar E_A,[\bar E_B,\bar E_C]}=\braket{\bar E_B,[\bar E_C,\bar E_A]}~.
\ee
Note that the generalized vielbein $E_A{}^N$ defined in (\ref{vielbein_drin}) contains the fluctuation $U_A{}^B\in O(1,D-1)\times O(D-1,1)\backslash O(D,D)$ around the basis $\bar E_A$. We parametrize $U_A{}^B$ as 
\be
U_A{}^B=
\begin{pmatrix}
e^{-T}&0\\
eB&e
\end{pmatrix}~,\label{U_matrix}
\ee
where $e\in GL(D)$ and $B$ is a $D$-dimensional antisymmetric matrix.

In order to derive a $D$-dimensional action from the DFT action ({\ref{action_PLTd}), we apply the parametrization of an element of $\mathcal{D}$ described in the previous section, $l(X)=\bar g(\bar x)g(x)$ or $l(X')=g(x')\bar g(\bar x')$.

\subsubsection*{Case 1: $l=\bar g(\bar x)g(x)$}
For the case $l=\bar g(\bar x)g(x)$ the left invariant current is given by
\be
l^{-1}dl=\bar E^{-1}{}_M{}^AT_AdX^M=g^{-1}\bar g^{-1}d\bar gg+g^{-1}dg~.
\ee
From this expression we can read off $\bar E^{-1} _M{}^B$ as
\be
\bar E^{-1} _M{}^B = \begin{pmatrix}\bar L^{-1}R^{-T}&\\
&1
\end{pmatrix}
\begin{pmatrix}
L^T&\\
&L^{-1}
\end{pmatrix}
\begin{pmatrix}
1&-\Pi\\
&1
\end{pmatrix}_M{}^B\ ,
\ee
where $ L$ is given in (\ref{left_invariant_of_g}), $\bar L$ is given in (\ref{left_invariant_of_gbar}), and the right invariant vector field $R$ of $\CG$ is defined by
\be
dgg^{-1}=R^{-1}_m{}^adx^mt_a~.
\ee
The generalized metric is given by 
\bea
H_{MN}&=& E^{-1}_M{}^AH_{AB} E^{-1B}{}_N = (\bar E^{-1} U^{-1} H U^{- T} \bar E^{- T} )_M{}_N\cr 
%&=&\begin{pmatrix}
%\bar L^{-1}R^{-T}&\\
%&1
%\end{pmatrix}
%\begin{pmatrix}
%L^T&\\
%&L^{-1}
%\end{pmatrix}
%\begin{pmatrix}
%1&-\Pi\\
%&1
%\end{pmatrix}
%\begin{pmatrix}
%1&\\
%-B&1
%\end{pmatrix}
%\begin{pmatrix}
%e^T&\\
%&e^{-1}
%\end{pmatrix}
%\begin{pmatrix}
%s^{-1}&\\
%&s
%\end{pmatrix}\cr
%&&\begin{pmatrix}
%e&\\
%&e^{-T}
%\end{pmatrix}
%\begin{pmatrix}
%1&B\\
%&1
%\end{pmatrix}
%\begin{pmatrix}
%1&\\
%\Pi&1
%\end{pmatrix}
%\begin{pmatrix}
%L&\\
%&L^{-T}
%\end{pmatrix}
%\begin{pmatrix}
%R^{-1}\bar L^{-T}&\\
%&1
%\end{pmatrix}\cr
&=&
\begin{pmatrix}
\bar L^{-1}R^{-T}&\\
&1
\end{pmatrix}
\begin{pmatrix}
L^T&\\
&L^{-1}
\end{pmatrix}
\begin{pmatrix}
1&-\Pi\\
&1
\end{pmatrix}
\begin{pmatrix}
G^{-1}&G^{-1}B\\
-BG^{-1}&G-BG^{-1}B
\end{pmatrix}\cr
&&\begin{pmatrix}
1&\\
\Pi&1
\end{pmatrix}
\begin{pmatrix}
L&\\
&L^{-T}
\end{pmatrix}
\begin{pmatrix}
R^{-1}\bar L^{-T}&\\
&1
\end{pmatrix}\cr
\cr&=&
\begin{pmatrix}
\bar L^{-1}R^{-T}&\\
&1
\end{pmatrix}
\begin{pmatrix}
\hat G^{-1}&\hat G^{-1}\hat B\\
-\hat B\hat G^{-1}&\hat G-\hat B\hat G^{-1}\hat B
\end{pmatrix}
\begin{pmatrix}
R^{-1}\bar L^{-T}&\\
&1
\end{pmatrix} \ .
%\cr&=&
%(\hat A^{-1}\hat H\hat A)_{MN}
\eea
%where we have introduced a matrix
%\be
%\hat A_M{}^N=
%\begin{pmatrix}
%R^T\bar L&\\
%&1
%\end{pmatrix}~,
%\ee
%\be
%\hat H_{AB}=\begin{pmatrix}
%\hat G^{-1}&\hat G^{-1}\hat B\\
%-\hat B\hat G^{-1}&\hat G-\hat B\hat G^{-1}\hat B
%\end{pmatrix}_{AB}~,\label{def_hatE_H}
%\ee
 Here, $\hat G$ and $\hat B$ are defined by
\be
\hat G_{mn}+\hat B_{mn}=L^{-1}{}_m{}^a\Big(\frac{1}{\frac{1}{G+B}+\Pi}\Big)_{ab}L^{-T}{}^b{}_n~.
\label{def_hat_GB}
\ee
It is convenient to introduce a new basis $\hat{E}_M$, called 'hat-basis' in the following, such that 
\bea
\hat{E}_N&:=&\hat A_N{}^M\partial_M
\cr
&=&\hat A_N{}^M\bar E_M{}^A\bar E_A\cr
&=&
\begin{pmatrix}
L^T&-L^T\Pi\\
&L^{-1}
\end{pmatrix}
{}_N{}^A\bar E_A~.
\label{relation_barn_bara1}
\eea
The relation between the local basis $E_A$ and the hat-basis $\hat E_M$ is given by
a matrix $\hat E_A{}^M\in O(D,D)$ as
\be
E_A=\hat E_A{}^M\hat E_M\label{def_hat_vierbein}.
\ee
\if0
\be
E_A{}^M=\hat E_A{}^N\hat A_N{}^M~,~
\hat A_M{}^N=
\begin{pmatrix}
R^T\bar L&\\
&1
\end{pmatrix}~.
\label{def_hat_vierbein}
\ee
\fi
Then, we also define $\hat H_{MN}$ and $\hat \eta_{MN}$ in this basis.
In terms of $\hat G$ and $\hat B$, they are given as
\bea
\hat H_{MN}&:=&\hat{E}^{-1}{}_M{}^AH_{AB}\hat{E}^{-T}{}^B{}_N=
\begin{pmatrix}
\hat G^{-1}&\hat G^{-1}\hat B\\
-\hat B\hat G^{-1}&\hat G-\hat B\hat G^{-1}\hat B
\end{pmatrix}~,\label{def_hatE_H}\\
\hat\eta_{MN}&:=&\hat{E}^{-1}{}_M{}^A\eta_{AB}\hat{E}^{-T}{}^B{}_N=
\begin{pmatrix}
0&1\\
1&0
\end{pmatrix}~.\label{def_hatE_eta}
\eea
The relation between these bases is therefore
\be
\partial_M\stackrel{\hat A_N{}^M}{\longmapsto}\hat{ E}_N\stackrel{\hat E_A{}^N}{\longmapsto}E_A~.
\ee
The flux in the basis $\hat { E}_M$ can be calculated by using the flux in $\bar{E}_A$ as follows.
Using (\ref{relation_barn_bara1}) we obtain 
\be
\hat{F}'_{LMN}:=\braket{[\hat{ E}_L,\hat{ E}_M]_L,\hat{ E}_N}~,
\ee
The non-zero components are
\bea
%\hat{ F}'{}_{lm}{}^n&=&0~,\cr
%\hat{ F}'{}_{lmn}&=&0~,\cr
\hat{ F}'{}_{l}{}^m{}_n&=&\partial_lL_c{}^mL^{-1}{}_n{}^c-L^{-1}{}_l{}^aL_b{}^mf_{ac}{}^bL^{-1}{}_n{}^c~,\cr
%\hat{ F}'{}_l{}^{mn}&=&0~,\cr
\hat{ F}'{}^{lm}{}_n&=&\bar f^{lm}{}_{n}+2f_{np}{}^{[l}\Pi^{m]p}~.
%\hat{ F}'{}^{lmn}&=&0~,\\
\eea
For $\hat\phi'$ given by
\be
\hat{{\phi}}'{}_{LMN}:=\braket{[\hat{ E}_L,\hat{ E}_M]-[\hat{ E}_L,\hat{ E}_M]_L,\hat{ E}_N}~,
\ee
the non-zero components are
%\hat{{\phi}}'{}_{lm}{}^n&=&0~,\cr
%\hat{{\phi}}'{}_{lmn}&=&0~,\cr
\bea
\hat{{\phi}}'{}_{l}{}^m{}_n&=&\partial_n L^{-1}{}_l{}^aL{}_a{}^m~,\cr
%\hat{{\phi}}'{}_l{}^{mn}&=&0~,\cr
\hat{{\phi}}'{}^{lm}{}_n&=&-(\bar f^{lm}{}_n+2f_{nn'}{}^{[l}\Pi^{m]n'})\cr
&=&-L_a{}^lL_b{}^m\partial_n\Pi^{ab}~.
%\hat{{\phi}}'{}^{lmn}&=&0~,\\
\eea
Then we obtain the relation
\be
\hat{{F}}_{LMN}:=\braket{[\hat{ E}_L,\hat{ E}_M],\hat{ E}_N}=\hat{{F}}'_{LMN}+\hat{{\phi}}'_{LMN}=0~.
\ee
For the flux with one index we get
\be
\hat{F}_M:=\hat{ \phi}'_{LM}{}^L+2\rho(\hat{ E}_M)(d)~,
\ee
and in components
\bea
\hat{F}_m&=&2\partial_m(d+\frac{1}{2}\log\det(R))~,\cr
\hat{ F}^{m}&=&-R_b{}^m\bar f^{ab}{}_a+2\rho(\hat E^m)(d)~.
\eea
Note that $\hat F_M$ has contributions 
from $\hat \phi'_{NM}{}^N$ besides the term $2\partial_M d$.\footnote{When the generalized dilaton depends on $x^m$ only and we consider the unimodular case,
$d+\frac{1}{2}\log\det(R)$ corresponds to the generalized dilaton in DFT$_{sec}$.}  
As we will show, the Poisson-Lie T-duality of the dilaton is automatically 
achieved due to these contributions from $\hat \phi'_{NM}{}^N$ without the use of a linear dilation.

We write the DFT action in the hat-basis using the field $\hat E_A{}^N$ given in (\ref{def_hat_vierbein}) as
\bea
F_{ABC}&=&3\hat{\Omega}_{[ABC]}+\hat{E}_A{}^L\hat{E}_B{}^M\hat{E}_C{}^N\hat{{F}}_{LMN}\cr
&=&3\hat{\Omega}_{[ABC]}~,\\
F_A&=&\hat{\Omega}^B{}_{BA}+\hat{E}_A{}^N\hat{{F}}_N~,\\
\phi'_{ABC}&=&\hat{\Omega}_{CAB}+\hat{E}_A{}^L\hat{E}_B{}^M\hat{E}_C{}^N\hat{{\phi}}'{}_{LMN}~.
\eea
$\hat\Omega_{ABC}$ is the Weizenb\"ock connection
\bea
\hat{\Omega}_{ABC}&=&\hat E_A{}^L\rho(\hat{E}_L)(\hat{E}_B{}^M)\hat{E}_C{}^N\hat{\eta}_{MN}~.
\eea
For the measure we obtain
\bea
c_0\int dXe^{-2d}\sqrt{\det\eta_{MN}}&=&c_0\int dX e^{-2d}\sqrt{(\det{\hat E^{-1}{}_L{}^A})^2(\det{\hat A^{-1}{}_M{}^N})^2}\cr
&=&c_0\int dX e^{-2d}\det R^{-1}\det \bar L^{-1}\cr
&=&c_0\int dXe^{-2(d+\frac{1}{2}\log\det R)}\det \bar L^{-1}~.
\eea
We denote the DFT action (\ref{action_PLTd}) with the measure and flux written in the hat-basis by $\hat S_{DFT}[U,d]$.

What we can immediately see is the following:
When $U$ and $d$ depend on $x^m$ only, the $\bar x_m$ integration factorizes (with the factor $\det \bar L^{-1}$) and $\hat S_{DFT}[U(x),d(x)]$ reduces to the D-dimensional action up to a normalization.
We find that the e.o.m. of the action $\hat S_{DFT}[U(x),d(x)]$ is, in general, the Generalized Supergravity Equation (GSE). Note that this procedure does not mean that we use the section condition, where it is allowed to introduce a linear dilaton depending on
 the dual coordinate. Here, we apply a dimensional reduction of the fluctuation and thus, there is no dual coordinate dependence.
 
Denoting a solution of the e.o.m. of the action $\hat S_{DFT}[U(x),d(x)]$ by $U_0(x), d_0(x)$, where $U_0(x)$ is determined by $G_0$ and $B_0$ via (\ref{U_matrix}), a solution of the GSE is given by
\bea
\hat G+\hat B&=&L^{-1}\frac{1}{\frac{1}{G_0+B_0}+\Pi}L^{-T}~,\\
\hat\phi&=&d_0+\frac{1}{2}\log\det R+\frac{1}{4}\log\det \hat G~,\\
\hat I^m&=&R_b{}^m(x)\bar f^{ab}{}_a~.
\eea
In the equations above $\hat \phi$ is the dilaton and $\hat I^m$ is a Killing vector in the GSE.

The easiest way to see this is as follows.
Let us denote the modified DFT action given in \cite{Sakatani_2017} as $
\hat S_{DFT_{sec}}^{mod}[E_A{}^N(x),d(x),\hat I^m(x)]$, for details see appendix B.
We find that $\hat S_{DFT_{sec}}^{mod}[E_A{}^N(x),d(x),\hat I^m(x)]$ and $\hat S[U(x),d(x)]$ are related as
\be
\hat S_{DFT}[U(x),d(x)]=\Big(\int d\bar x\det \bar L^{-1}\Big) \hat S_{DFT_{sec}}^{mod} [\hat E_A{}^N(x),d(x)+\frac{1}{2}\log\det R(x),R_b{}^m(x)\bar f^{ab}{}_a]~.\label{factorization_of_measure}
\ee
The $\bar x_m$ dependence in this action exists only in the factor $\int d\bar x\det \bar L^{-1}$.
Thus, when $U$ and $d$ depend only on $x^m$, 
$\hat S[U(x),d(x)]$ reduces to the GSE.

\subsubsection*{Case 2: $l=g(x')\bar g(\bar x')$}
Next, we consider the case where $l=g(x')\bar g(\bar x')$. This case is connected to the previous one by a coordinate transformation.
The generalized metric is
\bea
H_{MN}&=&E_M{}^AH_{AB}E^B{}_N\cr
&=&\begin{pmatrix}
1&\\
&L^{-1}\bar R^{-T}
\end{pmatrix}
\begin{pmatrix}
\bar {L}^{-1}&\\
&\bar L^{T}
\end{pmatrix}
\begin{pmatrix}
1&\\
-\bar\Pi&1
\end{pmatrix}
\begin{pmatrix}
1&\\
-B&1
\end{pmatrix}
\begin{pmatrix}
e^T&\\
&e^{-1}
\end{pmatrix}
\begin{pmatrix}
s^{-1}&\\
&s
\end{pmatrix}\cr
&&\begin{pmatrix}
e&\\
&e^{-T}
\end{pmatrix}
\begin{pmatrix}
1&B\\
&1
\end{pmatrix}
\begin{pmatrix}
1&\bar\Pi\\
&1
\end{pmatrix}
\begin{pmatrix}
\bar L^{-T}&\\
&\bar L
\end{pmatrix}
\begin{pmatrix}
1&\\
&\bar R^{-1}L^{-T}
\end{pmatrix}\cr
&=&
\begin{pmatrix}
1&\\
&L^{-1}\bar R^{-T}
\end{pmatrix}
\begin{pmatrix}
\check G-\check B\check G^{-1}\check B&-\check B\check G^{-1}\\
\check G^{-1}\check B&\check G^{-1}
\end{pmatrix}
\begin{pmatrix}
1&\\
&\bar R^{-1}L^{-T}
\end{pmatrix}~,
\eea
where $\check G^{mn}+\check B^{mn}$ is defined by
\be
\check G^{mn}+\check B^{mn}=\bar L^{-1}{}^m{}_a\Big(\frac{1}{G+B+\bar\Pi}\Big)^{ab}\bar L^{-T}{}_b{}^n~.
\ee
We separate $E_A{}^M$ into an $O(D,D)$ part $\check E_A{}^N$ and a part denoted by $\check A_N{}^M$.
\be
E_A{}^M=\check E_A{}^N\check A_N{}^M~,~
\check A_M{}^N=
\begin{pmatrix}
1&\\
&\bar{R}^TL
\end{pmatrix}~.
\ee
Similarly as before, we have now introduced the 'check-basis' $\check E_A{}^N$ such that
\be
\check E_N=\check{E}^{-1}{}_N{}^AE_A=\check A_N{}^M\partial_M=
\begin{pmatrix}
1&\\
&\bar{R}^TL
\end{pmatrix}{}_N{}^M\partial_M~.
\ee
and
\bea
\check E_N
&=&
\begin{pmatrix}
\bar L^{-1}&\\
-\bar L^T\bar\Pi&\bar L^T
\end{pmatrix}
{}_N{}^A\bar E_A~.
\label{relation_barn_bara}
\eea
The relation between the bases is
\be
\partial_M\stackrel{\check A_N{}^M}{\longmapsto}\check{ E}_N\stackrel{\check E_A{}^N}{\longmapsto}E_A~.
\ee
The metrics in the check-basis $\check E_M$ are given by
\bea
\check{H}_{MN}&=&\check{E}^{-1}{}_M{}^AH_{AB}\check{E}^{-T}{}^B{}_N=
\begin{pmatrix}
\check G-\check B\check G^{-1}\check B&-\check B\check G^{-1}\\
\check G^{-1}\check B&\check G^{-1}
\end{pmatrix}~,\\
\check{\eta}_{MN}&=&\check{E}^{-1}{}_M{}^A\eta_{AB}\check{E}^{-T}{}^B{}_N=
\begin{pmatrix}
0&1\\
1&0
\end{pmatrix}~.
\eea
The fluxes in the check-basis $\check {E}_M$ can be calculated from the corresponding quantities in the  $\bar E_A$ basis. They are related as:
\be
\check{F}'_{LMN}:=\braket{[\check{ E}_L,\check{ E}_M]_L,\check{ E}_N}~,
\ee
with non-zero components
%\check{F}'{}^{lm}{}_n&=&0~,\cr
%\check{F}'{}^{lmn}&=&0~,\cr
\bea
\check{F}'{}^{l}{}_m{}^n&=&\partial^l\bar L^c{}_m\bar L^{-1}{}^n{}_c-\bar L^{-1}{}^l{}_a\bar L^b{}_m\bar f^{ac}{}_b\bar L^{-1}{}^n{}_c~,\cr
%\check{F}'{}^l{}_{mn}&=&0~,\cr
\check{F}'{}_{lm}{}^n&=&f_{lm}{}^{n}+2\bar f^{np}{}_{[l}\bar \Pi_{m]p}~.
%\check{F}'{}_{lmn}&=&0~.\\
\eea
For $\check{{\phi}}'{}_{LMN}$
\be
\check{{\phi}}'{}_{LMN}:=\braket{[\check{ E}_L,\check{ E}_M]-[\check{ E}_L,\check{ E}_M]_L,\check{ E}_N}~,
\ee
with non-zero components
\bea
%\check{\phi}'{}^{lm}{}_n&=&0~,\cr
%\check{\phi}'{}^{lmn}&=&0~,\cr
\check{\phi}'{}^{l}{}_m{}^n&=&\partial^n \bar L^{-1}{}^l{}_a\bar L{}^a{}_m~,\cr
%\check{\phi}'{}^l{}_{mn}&=&0~,\cr
\check{\phi}'{}_{lm}{}^n&=&-(f_{lm}{}^n+2\bar f^{nn'}{}_{[l}\bar \Pi_{m]n'})\cr
&=&-\bar L^a{}_l\bar L^b{}_m\partial^n\bar\Pi_{ab}~.
%\check{\phi}'{}_{lmn}&=&0~.\\
\eea
Thus, we get the relation
\be
\check{{F}}_{LMN}:=\braket{[\check{ E}_L,\check{ E}_M],\check{ E}_N}=\check{{F}}'_{LMN}+\check{{\phi}}'_{LMN}=0~.
\ee
For the flux with one index
\bea
\check{F}_M&:=&\check{ \phi}'_{LM}{}^L+2\rho(\check{ E}_M)(d)~,\cr
\check F^m&=&2\partial^m(d+\frac{1}{2}\log\det(\bar R))~,\cr
\check F_{m}&=&-\bar R^b{}_m f_{ab}{}^a+2\rho(\check E_m)(d)~.
\eea
To obtain the DFT action in the check-basis, we rewrite the fluxes as
\bea
F_{ABC}&=&3\check{\Omega}_{[ABC]}+\check{E}_A{}^L\check{E}_B{}^M\check{E}_C{}^N\check{F}_{LMN}\cr
&=&3\check{\Omega}_{[ABC]}~,\\
F_A&=&\check{\Omega}^B{}_{BA}+\check{E}_A{}^N\check{F}_N~,\\
\phi'_{ABC}&=&\check{\Omega}_{CAB}+\check{E}_A{}^L\check{E}_B{}^M\check{E}_C{}^N\check{\phi}'{}_{LMN}~,
\eea
where $\check\Omega_{ABC}$ is the corresponding Weizenb\"ock connection
\bea
\check{\Omega}_{ABC}&=&\check E_A{}^L\rho(\check{E}_L)(\check{E}_B{}^M)\check{E}_C{}^N\check{\eta}_{MN}~.
\eea
The measure of the DFT action in the original basis is related to the one in the check-basis as
\be
c_0\int dXe^{-2d}\sqrt{\det{\eta}_{MN}}=c_0\int dXe^{-2(d+\frac{1}{2}\log\det \bar R)}\det L^{-1}~.
\ee
Similar to the previous case, we denote the DFT action in the check-basis by $\check S_{DFT}[U,d]$.
The relation between $\check S_{DFT_{sec}}^{mod}[E_A{}^N(\bar x'),d(\bar x'),\check I^m(\bar x')]$ generating 
the GSE and $\check S_{DFT}[U(\bar x'),d(\bar x')]$ is
\be
\check S_{DFT}[U(\bar x'),d(\bar x')]=\Big(\int d x'\det  L^{-1}\Big)\check S_{DFT_{sec}}^{mod}[\check E_A{}^N(\bar x'),d(\bar x')+\frac{1}{2}\log\det \bar R(\bar x'),\bar R^b{}_m(\bar x')f_{ab}{}^a]~.
\ee
On the r.h.s. only the integration measure depends on $x'{}^m$ giving an overall factor $\int d x'\det  L^{-1}$.
When $U$ and $d$ depend only on $\bar x'_m$, the action 
$\check S_{DFT}[U(\bar x'),d(\bar x')]$ reduces to the GSE.

\subsection{SUGRA solution and DFT solution}
\label{sec:SUGRAvsDFT}
In the previous section, we have assumed that $U_0(x)$ and $d_0(x)$ are solutions of the e.o.m. of the D-dimensional action $\hat S[U(x),d(x)]$, i.e., 
\be
\hat S[U_0(x)+\delta U(x),d_0(x)+\delta d(x)]-\hat S[U_0(x),d_0(x)]=0~.
\label{hatS_fluctuation_x}
\ee
This means that there exist the e.o.m. for $U_{AB}(x)$ and $d(x)$ denoted by ${\sf C}_{AB}(x,U_0,d_0),{\sf C}(x,U_0,d_0)$ such that,
\bea
0&=&\hat S[U_0(x)+\delta U(x),d_0(x)+\delta d(x)]-\hat S[U_0(x),d_0(x)]\cr
&=&\int dX\Big(\delta U^{AB}(x){\sf C}_{AB}(x,U_0,d_0)+\delta d(x){\sf C}(x,U_0,d_0)\Big)~.
\eea
Now, we extend the fluctuation over the full doubled space $X$ and consider 
\be
\hat S[U_0(x)+\delta U(X),d_0(x)+\delta d(X)]-\hat S[U_0(x),d_0(x)]~.
\ee
Note that $\delta U(X)$ and $\delta d(X)$ depend on all coordinates in general.
Compared to (\ref{hatS_fluctuation_x}), this variation has new contributions denoted as ${\sf C}_{mAB}(x,U_0,d_0), {\sf C}_{m}(x,U_0,d_0) \in C^\infty(M)$ via the Weizenb\"ock connnection $\hat\Omega_{ABC}$ and the flux $F_A$, and can be written as
\bea
&&\hat S[U_0+\delta U(X),d_0+\delta d(X)]-\hat S[U_0,d_0]\cr
&=&\int dX\Big(\delta U^{AB}(X){\sf C}_{AB}(x,U_0,d_0)+\delta d(X){\sf C}(x,U_0,d_0)\Big)\cr
&&+c_0\int dXe^{-2d_0}\sqrt{\det\eta_{MN}}\Big(\rho(\hat E^m)(\delta U^{AB}(X)){\sf C}_{mAB}(x,U_0,d_0)+\rho(\hat E^m)(\delta d(X)){\sf C}_{m}(x,U_0,d_0) \Big)\cr
&=&c_0\int dXe^{-2d_0}\sqrt{\det\eta_{MN}}\Big(\delta U^{AB}(X)(-\rho(\hat E^m)+\hat F^m(d_0))({\sf C}_{mAB}(x,U_0,d_0))\cr
&&+\delta d(X)(-\rho(\hat E^m)+\hat F^m(d_0))({\sf C}_{m}(x,U_0,d_0)) \Big)\cr
&=&c_0\int dXe^{-2d_0}\sqrt{\det\eta_{MN}}\Big(\delta U^{AB}(X)\hat F^m(d_0){\sf C}_{mAB}(x,U_0,d_0)\cr
&&+\delta d(X)\hat F^m(d_0){\sf C}_{m}(x,U_0,d_0) \Big)~.
\eea
Since $d_0$ depends only on $x^m$, $\hat F^m(d_0)$ is given by
\be
\hat F^m(d_0)=-R_b{}^m\bar f^{ab}{}_a~.
\ee
Thus, if $\bar{\mathcal{G}}$ is unimodular, $\bar f^{ab}{}_a=0$, 
then $U_0$ and $d_0$ are solutions of the $2D$-dimensional DFT, i.e.,
\be
\hat S[U_0+\delta U(X),d_0+\delta d(X)]-\hat S[U_0,d_0]=0~.
\ee

Now, we are ready to derive the Poisson-Lie T-duality using the above considerations as follows.
\begin{enumerate}
\item Given a solution of the GSE
\bea
\hat G+\hat B&=&L^{-1}\frac{1}{\frac{1}{G_0+B_0}+\Pi}L^{-T}~,\\
\hat\phi&=&d+\frac{1}{2}\log\det R+\frac{1}{4}\log\det \hat G~,\\
\hat I^m&=&R_b{}^m(x)\bar f^{ab}{}_a~,
\eea
where $G_0$ and $B_0$ depend only on $x^m$.
This means that $U_0$ and $d_0$ satisfy
\be
\hat S[U_0(x)+\delta U(x),d_0(x)+\delta d(x)]-\hat S[U_0(x),d_0(x)]=0~.
\ee

\item	If $\bar{\mathcal{G}}$ is unimodular $(\bar f^{ab}{}_a=0)$, $U_0$ and $d_0$ are a solution of the $2D$-dimensional DFT action, i.e.,
\be
\hat S[U_0(x)+\delta U(X),d_0(x)+\delta d(X)]-\hat S[U_0(x),d_0(x)]=0~.
\label{NSNSsector_full_fluctuation}
\ee
Note that in this case $\hat I^m=0$, meaning that $(\hat G,\hat B,\hat \phi)$ is a solution for SUGRA.
\item	Then, we perform a coordinate transformation  such that $\bar g(\bar x)g(x)=g(x')\bar g(\bar x')$,
we obtain
\be
\check S[U_0(x(X'))+\delta U(X'),d_0(x(X'))+\delta d(X')]-\check S[U_0(x(X')),d_0(x(X'))]=0~.
\ee
\item	Since we performed the coordinate transformation of the fluctuation in full space,
it holds also in the restricted space, $\delta U(\bar x'_m),\delta d(\bar x'_m)$,
 %and the action is still invariant under the fluctuations $\delta U$ and $\delta d$, i.e.,
\be
\check S[U_0(x(X'))+\delta U(\bar x'),d_0(x(X'))+\delta d(\bar x')]-\check S[U_0(x(X')),d_0(x(X'))]=0~.
\label{P-L_dual_eom}
\ee

\item	In general, $U_0(x(X'))$ depends on both ${\bar x}'{}^m$ and $x'{}^m$.
However, in such a case, $U_0$ can not generate a solution on a $D$-dimensional space. 
Therefore, in order to obtain $\check G$ and $\check B$ as $D$-dimensional background, 
we require that $U_0(x(X'))$ depends only on $\bar x'_m$.
\be
\frac{\partial}{\partial x'{}^m} U_0(x(X'))=0~.
\label{cond_D-dimensional_U0}
\ee

\item	On the other hand, the condition for $d_0$ can be relaxed, 
since the $x'{}^m$-dependence of $d_0$ can vanish by a redefinition of $d_0$ and $\check I_m$, 
as shown in the following. 
We require
\bea
d_0(x(X'))&=&d_1(x')+d_2(\bar x')~,\label{cond_d_0}\\
\frac{\partial}{\partial x'{}^n}(\rho(\check E_m)(d_1))&=&0~.\label{cond_d_02}
\label{P-L_cond_dilaton}
\eea
Then, we can define $d'_0$ and $\check I_m'$ which depend only on $\bar x'_m$.
Using $\check E_m=\bar R^T_m{}^aL_a{}^l\partial_l$, the condition (\ref{P-L_cond_dilaton}) means
\be
L_a{}^l\partial_ld_1=J_a~,~(J_a:\mbox{constant})~.
\ee
$d'_0$ and $\check I_m'$ are given by
\bea
d'_0(\bar x')&=&d_2(\bar x')~, \label{cond3} \\
\check I_m'(\bar x')&=&\check I_m(\bar x')-2\rho(\check E_m)(d_1)\cr
&=&\check I_m(\bar x')-2\bar R_m^T{}^a(\bar x')J_a~.\label{cond4}
\eea
$d_0'$ and $\check I_m'$ do not depend on $x'$.\footnote{Note that originally $U_0$ and $d_0$ are assumed to be constant, meaning that the conditions given in (\ref{cond_D-dimensional_U0}), (\ref{cond_d_0}) and (\ref{cond_d_02}) are satisfied. Then it follows that the shift of  $I_m$ in (\ref{condIprime}) vanishes.}

\item	Finally, we obtain a solution of the GSE in the dual space.
\bea
\check G^{mn}+\check B^{mn}&=&\bar L^{-1}\frac{1}{G_0+B_0+\bar \Pi}\bar L^{-T}~,\\
\check \phi&=&d'_0+\frac{1}{2}\log\det \bar R+\frac{1}{4}\log\det \check G\cr
&=&d_2+\frac{1}{2}\log\det \bar R+\frac{1}{4}\log\det \check G~,\\
\check I'_m&=&\bar R^b{}_m f_{ab}{}^a-2\rho(\check E_m)(d_1)~. \label{condIprime}
\eea

\end{enumerate}

In the following we show how the redefinition in step 6 is obtained. 
First, we consider a redefinition such that the flux $\check F_M$ keeps its form as
\bea
\check F^m&=&2\partial^m(d_0(x(X'))+\frac{1}{2}\log\det(\bar R))=2\partial^m(d'_0+\frac{1}{2}\log\det(\bar R))~,\\
\check F_m&=&-\check I_m(\bar x')+2\rho(\check E_m)(d_0(x(X')))=-\check I'_m+2\rho(\check E_m)(d'_0)~.
\eea
To obtain a $D$-dimensional solution, 
$d_0'$ and $\check I'_m$ have to depend on $\bar x_m'$ only, i.e.,
\bea
\check F^m&=&2\partial^m(d_0(x(X'))+\frac{1}{2}\log\det(\bar R))=2\partial^m(d'_0(\bar x')+\frac{1}{2}\log\det(\bar R))~,
\label{cond_redefF^m}
\\
\check F_m&=&-\check I_m(\bar x')+2\rho(\check E_m)(d_0(x(X')))=-\check I'_m(\bar x')~.
\label{cond_redefF_m}
\eea
Considering that $\partial'_n$ acts on both sides, 
we get the necessary condition,
\bea
2\partial'_n(\partial'{}^md_0(x(X')))&=&0~,\\
2\partial'_n(\rho(\check E_m)(d_0(x(X'))))&=&0~.
\eea
Thus, the conditions (\ref{cond_d_0}), (\ref{cond_d_02}) for $d_0(x(X'))$ are sufficient.
%\bea
%d_0(x(X'))&=&d_1(x')+d_2(\bar x')~,\\
%\frac{\partial}{\partial x'{}^n}(\rho(\check E_m)(d_1))&=&0~.
%\eea
From (\ref{cond_redefF^m}) and (\ref{cond_redefF_m}) we obtain $d_0'$ and $\check I'_m$ given in (\ref{cond3}) and (\ref{cond4}), respectively.
%\bea
%d'_0(\bar x')&=&d_2(\bar x')~,
%\label{cond_D-dimensional_d0}\\
%\check I_m'(\bar x')&=&\check I_m(\bar x')-2\rho(\check E_m)(d_1)~.
%\label{cond_D-dimensional_I}
%\eea

The redefinition $d'=d-d_1,\check I_m'=\check I_m-2\rho(\check E_m)(d_1)$ changes the measure of the action as
\be
c_0\int dX\sqrt{\det \eta_{MN}}e^{-2d}=c_0\int dX\sqrt{\det \eta_{MN}}e^{-2d'}e^{-2d_1}~.
\ee
Thus, the relation between $\check S$ and $\check S_{DFT_{sec}}^{mod}$ is given by
\bea
\check S[U(X'),d(X')]&=&\Big(\int d x'\det  L^{-1}\Big)\check S_{DFT_{sec}}^{mod}[\check E_A{}^N(X'),d(X(X'))+\frac{1}{2}\log\det \bar R(\bar x'),\bar R^b{}_m(\bar x')f_{ab}{}^a]\cr
&=&\Big(\int d x'\det  L^{-1}e^{-2d_1}\Big)\cr
&&\times \check S_{DFT_{sec}}^{mod}[\check E_A{}^N(X'),d_2+\frac{1}{2}\log\det \bar R,\bar R^b{}_mf_{ab}{}^a-2\rho(\check E_m)(d_1)]~.
\label{factrized_x-dependence}
\eea
Using these equations, (\ref{P-L_dual_eom}) becomes
\bea
0&=&\check S[U_0(x(X'))+\delta U(\bar x'),d_0(x(X'))+\delta d(\bar x')]-\check S[U_0(x(X')),d_0(x(X'))]\cr
&=&\Big(\int d x'\det  L^{-1}e^{-2d_1}\Big)\Big(\check S_{DFT_{sec}}^{mod}[\check E_0{}_A{}^N+\delta \check E_A{}^N,d'_0+\delta d,\check I'_m]-\check S_{DFT_{sec}}^{mod}[\check E_0{}_A{}^N,d'_0,\check I'_m]\Big)~,
\cr&&
\eea
where $\check E_0{}_A{}^N(\bar x')$ is defined by
\bea
\check{E}_0^{-1}{}_M{}^AH_{AB}\check{E}_0^{-T}{}^B{}_N&=&
\begin{pmatrix}
\check G_0-\check B_0\check G_0^{-1}\check B_0&-\check B_0\check G_0^{-1}\\
\check G^{-1}_0\check B_0&\check G^{-1}_0
\end{pmatrix}~,\\
\check G_0+\check B_0&=&\bar L^{-1}\frac{1}{G_0+B_0+\bar \Pi}\bar L^{-T}~,
\eea
Using (\ref{cond_D-dimensional_U0}), (\ref{cond3}) and (\ref{cond4}), we find that
$(\check E_0{}_A{}^N,d'_0,\check I'_m)$ is a $D$-dimensional solution generating 
a solution for the GSE.

We summarize the Poisson-Lie T-duality derived in this section:

%\begin{itembox}

\paragraph{Poisson-Lie T-duality in DFT}

We consider a solution for SUGRA given by
\bea
\hat G+\hat B&=&L^{-1}\frac{1}{\frac{1}{G_0+B_0}+\Pi}L^{-T}~,\\
\hat\phi&=&d_0+\frac{1}{2}\log\det R+\frac{1}{4}\log\det \hat G~,
\eea
where $\bar{\mathcal{G}}$ is unimodular, and $G_0$, $B_0$ and $d_0$ satisfy
\bea
\bar g(\bar x) g(x)&=&g(x')\bar g(\bar x')~,\\
\frac{\partial}{\partial x'{}^m}(G_0(x(X'))+B_0(x(X')))&=&0~,\\
d_0(x(X'))&=&d_1(x')+d_2(\bar x')~,\\
\frac{\partial}{\partial x'{}^n}(L_a{}^m\partial_m(d_1))&=&0~.
\eea
In this case, another solution of the GSE is constructed by
\bea
\check G^{mn}+\check B^{mn}&=&\bar L^{-1}\frac{1}{G_0+B_0+\bar \Pi}\bar L^{-T}~,
\label{check_E}\\
\check \phi&=&d_2+\frac{1}{2}\log\det \bar R+\frac{1}{4}\log\det \check G~,
\label{check_phi}\\
\check I'_m&=&\bar R^b{}_m f_{ab}{}^a-2\rho(\check E_m)(d_1)~.
\label{check_I'}
\eea

For constant $d_0$,
(\ref{check_phi}) and (\ref{check_I'}) simplify to
\bea
\check \phi&=&d_0+\frac{1}{2}\log\det \bar R+\frac{1}{4}\log\det \check G~,\\
\check I'_m&=&\bar R^b{}_m f_{ab}{}^a~.
\eea
In this case, we obtain a compact formula of the transformation rule for the dilaton without $d_0$ as
\be
e^{-2\hat\phi}=e^{-2\check\phi}\frac{\det(1+(G_0+B_0)\Pi)}{\det(1+(G_0+B_0)\bar\Pi)}\det a^{-1}~.
\ee
This T-duality of the dilaton is consistent with 
\cite{Tyurin_1996,Unge:2002aa,JURCO20181}, (See also \cite{Sakatani_2019}).

Since our discussion starts with a SUGRA solution,
we require that $\mathcal{G}$ is unimodular to get a corresponding solution for DFT.
On the other hand, as in the usual discussion, we can also start with a DFT solution, then
we obtain the T-duality where $\mathcal{G}$ and $\bar{\mathcal{G}}$ are not necessarily unimodular. In that case, $\bar a$ defined in (\ref{def_bar_a_Pi}) is not equal to $1$ in general,
and the transformation formula is given by
\be
e^{-2\hat\phi}=e^{-2\check\phi}\frac{\det(1+(G_0+B_0)\Pi)}{\det(1+(G_0+B_0)\bar\Pi)}\frac{\det \bar{a}}{\det a}~.
\ee
Note that for a given solution of GSE, a solution of DFT may not exist
as discussed in sec. \ref{sec:SUGRAvsDFT}.

\section{R-R sector}
The fields of the R-R sector can be assembled into a spinor $\chi$ \cite{Hohm:2011aa, Jeon:2012aa}. See also \cite{Fukuma:1999jt}.
To construct an action of the R-R sector, we define in the appendix a product called AK-product $(-,-)_{AK}$ for the spinors $\chi_1,\chi_2\in \mathbb{S}$ as 
\be
(\chi_1,\chi_2)_{AK}=\chi_1^\dagger A_+K\chi_2~.
\ee
The AK-product is invariant under an infinitesimal $O(1,D-1)\times O(D-1,1)$ transformation, 
i.e., for $\Lambda_{AB}=-\Lambda_{BA},\ \Lambda_{AB}=-H_B{}^{B'}H_A{}^{A'}\Lambda_{B'A'}$ we obtain
\be
(\chi_1,\chi_2)_{AK}={\Big(} \exp(\frac{1}{4}\Lambda_{AB}\gamma^{AB})\chi_1,\exp(\frac{1}{4}\Lambda_{A'B'}\gamma^{A'B'})\chi_2{\Big )}_{AK}\ .
\ee
The DFT action for the R-R sector $S_{RR}$ can be defined by
\be
S_{RR}=\beta_{RR}(\DGO\chi,\DGO\chi)_{AK}~,
\label{R-Rsector_action}
\ee
where $\chi\in \mathbb{S}$ is the R-R potential of DFT and $\beta_{RR}$ is a constant.
The constant $\beta_{RR}$ is determined by reducing $S_{RR}$ to the case of a flat background, and then comparing with the action in DFT${}_{sec}$. The result is  
\be
\beta_{RR}=-\frac{1}{2c_0}~.
\ee
For details, see appendix \ref{reductiontoDFT}. 

Note that a self-duality condition must be imposed on $\chi$ at the level of the equation of motion.
Usually, the R-R field is defined in the local coordinate.
Here, we construct the R-R field in the local Lorentz frame $E_A$ in the similar way as for the NS-NS sector.

\subsection{$O(1,D-1)\times O(D-1,1)$ transformation of R-R sector}
An $O(1,D-1)\times O(D-1,1)$ transformation of the R-R flux $F:=\DGO\chi$ is given by
\be
\chi\mapsto \SS_O\chi~,\label{chi_SOchi}
\ee
with $\SS_O$ being defined by 
\be
\SS_O\gamma_A\SS_O^{-1}=O_A{}^B\gamma_B~,
\ee
where the transformation $O_A{}^B$ is $O(1,D-1)\times O(D-1,1):
$\bea
O_A{}^{A'}O_B{}^{B'}\eta_{A'B'}&=&\eta_{AB}~,\\
O_A{}^{A'}O_B{}^{B'}H_{A'B'}&=&H_{AB}~.
\eea
The spin representation $\SS_O$ used in this paper satisfies
\bea
(\SS_O^{-1}{})^\dagger A_+K\SS_O^{-1}&=&(-1)^{\# O} A_+K~,\\
(\SS_O^{-1}K)^\dagger A_+K(\SS_O^{-1}K)&=&-(-1)^{\# O}A_+K~,
\eea
where $\# O$ is defined as the number of the transformations $(\Gamma_0\mapsto -\Gamma_0,\Gamma_{\bar 0}\mapsto \Gamma_{\bar 0})$ and $(\Gamma_0\mapsto \Gamma_0,\Gamma_{\bar 0}\mapsto -\Gamma_{\bar 0})$ in $O$.
Therefore, together with (\ref{chi_SOchi}) under the $O(1,D-1)\times O(D-1,1)$ transformation the fields transform as follows:
\be
E'_{A}{}^N= O_A{}^BE_B{}^N~,~d'= d~,~F'= \SS_O^{-1}F~.
\ee

The action of the NS-NS sector is invariant under this transformation.
On the other hand, the action $S_{RR}$ of the R-R sector is not invariant in general, i.e.,
\be
S'_{RR}= \beta_{RR}(\SS_O^{-1}F,\SS_O^{-1}F)_{AK}=(-1)^{\# O}S_{RR}~.
\ee
When $\# O=odd$, the $O(1,D-1)\times O(D-1,1)$ transformation is not a symmetry of DFT.
On the other hand, we can consider a local transformation 
\be
E''_{A}{}^N =O_A{}^BE_B{}^N~,~d''= d~,~F''= \SS_O^{-1}KF~.
\ee 
for $\# O=odd$.  With this transformation the DFT action is invariant, i.e.,
\be
S''_{RR}= \beta_{RR}(\SS_O^{-1}KF,\SS_O^{-1}KF)_{AK}=-(-1)^{\# O}S_{RR}=S_{RR}~.
\ee
Therefore, there exists a symmetry for $^\forall O\in O(1,D-1)\times O(D-1,1)$:
\be
(E_{A}{}^N~,~d~,~F)\mapsto
\begin{cases}
( O_A{}^BE_B{}^N~, d~,~\SS_O^{-1}F)~,&\# O=even~,\\
( O_A{}^BE_B{}^N~, d~,~\SS_O^{-1}KF)~,&\# O=odd~.
\end{cases}
\ee
This invariance guarantees the $O(1,D-1)\times O(D-1,1)$ covariance of the equation of motion after the 
self-duality condition.

To see this mechanism, we consider an $O(1,D-1)\times O(D-1,1)$ transformation of a solution of DFT given by
\be
E_A{}^N=E_{0A}{}^N~,~d=d_0~,~F=F_0~.
\ee
From the above discussion we can rotate solutions, i.e.,
\bea
E_A{}^N&=&O_A{}^BE_{0B}{}^N~,~d=d_0~,~F=\SS_{O}^{-1}F_0~,~(\# O:even)~,\label{soleven}\\
E_A{}^N&=&O_A{}^BE_{0B}{}^N~,~d=d_0~,~F=\SS_{O}^{-1}KF_0~,~(\# O:odd)~.\label{solodd}
\eea
However, for the e.o.m. we require the self-duality for the R-R field $F$ 
\be
KF=F~,
\ee
meaning that $\# O$ does not change the form of the solution and therefore (\ref{soleven}) and (\ref{solodd}) 
become the same equation
\bea
E_A{}^N&=&O_A{}^BE_{0B}{}^N~,~d=d_0~,~F=\SS_{O}^{-1}F_0~.
\label{ODD_map_into_solution}
\eea
This shows that the $O(1,D-1)\times O(D-1,1)$ transformation is not a symmetry of the action, in general,
but the e.o.m. is covariant and gives a mapping of a solution into another solution.

\subsection{Poisson-Lie T-duality of R-R sector}
In the previous section we have derived the Poisson-Lie T-duality of the NS-NS sector.
Here, we show the Poisson-Lie T-duality of the R-R sector.

Similar as in the discussion of NS-NS sector, we consider the case where $l=\bar g(\bar x) g(x)$,
\bea
\hat{E}_A{}^N=
\begin{pmatrix}
e^{-T}&0\\
0&e
\end{pmatrix}
\begin{pmatrix}
1&\\
B&1
\end{pmatrix}
\begin{pmatrix}
1&\Pi\\
0&1
\end{pmatrix}
\begin{pmatrix}
L^{-T}&0\\
0&L
\end{pmatrix}~.
\eea
The vielbein $\hat E_A{}^N$ generates 
the D-dimensional metric and B-field (\ref{def_hatE_H}).
Then, using a representation $^\exists \hat C\in O(1,D-1)\times O(D-1,1)$,
we write $\hat E_A{}^N$ as
\bea
\hat E_A{}^N=\hat C
\begin{pmatrix}
\hat e^{-T}&0\\
0&\hat e
\end{pmatrix}
\begin{pmatrix}
1&\\
\hat B&1
\end{pmatrix}~.
\eea
As in the previous section, the $GL(D)$ part and the B-transformation part are denoted by
\bea
\hat E^{(e)}_A{}^N&=&
\begin{pmatrix}
\hat e^{-T}{}^a{}_n&0\\
0&\hat e_a{}^n
\end{pmatrix}~,\\
\hat E^{(B)}_M{}^N&=&
\begin{pmatrix}
1^m{}_n&\\
\hat B_{mn}&1_m{}^n
\end{pmatrix}~.
\eea
To obtain the D-dimensional action, we define a rotated R-R flux $\hat F$ as
\be
\hat F=\sum_{p}\frac{1}{\sqrt{2^{p+1}}p!}\hat F_{m_1\cdots m_p}\delta_{a_1}^{m_1}\cdots \delta_{a_p}^{m_p}\gamma^{a_1\cdots a_p}\Ket{0}=e^{-\hat d}\SS_{\hat E^{(e)}}\DGO\chi~,
\ee
where $\hat d$ is defined by
\be
\hat d=d+\frac{1}{2}\log\det R~.
\ee
Using this $\hat F$, the action of the R-R sector is given by
\bea
S_{RR}&=&\beta_{RR}(\DGO\chi,\DGO\chi)_{AK}\cr
&=&\beta_{RR}{\Big (}e^{\hat d}\SS_{\hat E^{(e)}}^{-1}\hat F,e^{\hat d}\SS_{\hat E^{(e)}}^{-1}\hat F{\Big )}_{AK}\cr
&=&\beta_{RR}\sum_{p,q}{\Big (}e^{\hat d}\SS_{\hat E^{(e)}}^{-1}\frac{1}{\sqrt{2^{p+1}p!}}\hat F_{m_1\cdots m_p}\delta_{a_1}^{m_1}\cdots\delta_{a_p}^{m_p}\gamma^{a_1\cdots a_p}\Ket{0},\cr
&&~~~~~~~~~~~~~~e^{\hat d}\SS_{\hat E^{(e)}}^{-1}\frac{1}{\sqrt{2^{q+1}q!}}\hat F_{n_1\cdots n_q}\delta_{b_1}^{n_1}\cdots\delta_{b_q}^{n_q}\gamma^{b_1\cdots b_q}\Ket{0}{\Big )}_{AK}\cr
&=&\beta_{RR}\frac{1}{\sqrt{2^{p+1}p!}}\frac{1}{\sqrt{2^{q+1}q!}}\cr
&&\times\sum_{p,q}{\Big (}\SS_{\hat E^{(e)}}^{-1}\hat F_{m_1\cdots m_n}\delta_{a_1}^{m_1}\cdots\delta_{a_p}^{m_p}\gamma^{a_1\cdots a_p}\Ket{0},e^{2\hat d}\SS_{\hat E^{(e)}}^{-1}\hat F_{n_1\cdots n_q}\delta_{b_1}^{n_1}\cdots\delta_{b_q}^{n_q}\gamma^{b_1\cdots b_q}\Ket{0}{\Big )}_{AK}\cr
&=&\frac{1}{2}\beta_{RR}{\big (}\SS_{\hat E^{(e)}}^{-1}\Ket{0},\SS_{\hat E^{(e)}}^{-1}\Ket{0}{\big )}_{AK}e^{2\hat d}\hat G^{m_1n_1}\cdots \hat G^{m_pn_p}\hat F_{m_1\cdots m_p}\hat F_{n_1\cdots n_p}\cr
&=&\frac{1}{2}\beta_{RR}{\big (}\Ket{0},\Ket{0}{\big )}_{AK}\sqrt{\det \hat G_{ll'}}e^{2\hat d}\hat G^{m_1n_1}\cdots \hat G^{m_pn_p}\hat F_{m_1\cdots m_p}\hat F_{n_1\cdots n_p}\cr
&=&\frac{1}{2}\beta_{RR}c_0\int dX\sqrt{\det\eta_{MN}}e^{-2d}\sqrt{\det \hat G_{ll'}}e^{2\hat d}\hat G^{m_1n_1}\cdots \hat G^{m_pn_p}\hat F_{m_1\cdots m_p}\hat F_{n_1\cdots n_p}\cr
&=&-\frac{1}{4}\int dX\det R^{-1}\det \bar{L}^{-1}e^{-2d}\sqrt{\det \hat G_{ll'}}e^{2\hat d}\hat G^{m_1n_1}\cdots \hat G^{m_pn_p}\hat F_{m_1\cdots m_p}\hat F_{n_1\cdots n_p}\cr
&=&-\frac{1}{4}\int dX\det \bar{L}^{-1}\sqrt{\det \hat G_{ll'}}\hat G^{m_1n_1}\cdots \hat G^{m_pn_p}\hat F_{m_1\cdots m_p}\hat F_{n_1\cdots n_p}~.
\eea
When $U,d$ and $\hat F_{m_1\cdots m_p}$ depend only on $x^n$, the 
$\bar x_n$ integration simply gives a factor $\det \bar{L}^{-1}$ 
as in (\ref{factorization_of_measure}).

Let $F^{(0)}(x)$ be a solution of the e.o.m. for the action $S_{RR}$:
\be
\DGO\chi=F^{(0)}(x)~.
\ee
Note that $F_0$ depends only on $x^m$, 
and the D-dimensional R-R flux $\hat F$ is given by
\bea
\sum_{p}\frac{1}{\sqrt{2^{p+1}}p!}\hat F_{n_1\cdots n_p}\delta_{a_1}^{n_1}\cdots \delta_{a_p}^{n_p}\gamma^{a_1\cdots a_p}\Ket{0}&=&e^{-\hat d}\SS_{\hat E^{(e)}}F^{(0)}~.
\eea
With $^\exists \hat C_0\in O(1,D-1)\times O(D-1,1)$, 
the solution of the vielbein $\hat E_0{}_A{}^N$ is given by
\bea
E_0{}_{A}{}^N=
U_0
\begin{pmatrix}
1&\Pi\\
0&1
\end{pmatrix}
\begin{pmatrix}
L^{-T}&0\\
0&L
\end{pmatrix}=
\hat C_0
\begin{pmatrix}
\hat e^{-T}&0\\
0&\hat e
\end{pmatrix}
\begin{pmatrix}
1&\\
\hat B&1
\end{pmatrix}
~.
\eea
Then, we can write the R-R flux $\hat F$ by the algebraic structures $L,\Pi$ as
\bea
\sum_{p}\frac{1}{\sqrt{2^{p+1}}p!}\hat F_{n_1\cdots n_p}\delta_{a_1}^{n_1}\cdots \delta_{a_p}^{n_p}\gamma^{a_1\cdots a_p}\Ket{0}&=&e^{-\hat d}\SS_{\hat E^{(B)}}^{-1}\SS_{ L}\SS_{\Pi}\SS_{U_0}\SS_{\hat C_0}^{-1}F^{(0)}~,
\label{hat_F_solution_beforeODOD}
\eea
where the spin operators $\SS_L,\SS_\Pi,\SS_{U_0}$ and $\SS_{\hat C_0}$ are defined by
\bea
\SS_L\gamma_A\SS_L^{-1}&=&
\begin{pmatrix}
L^{-T}{}^a{}_m&\\
&L_a{}^m
\end{pmatrix}
\delta_M{}^B\gamma_B~,\\
\SS_\Pi\gamma_A\SS_\Pi^{-1}&=&
\begin{pmatrix}
1^a{}_b&\Pi^{ab}\\
&1_a{}^b
\end{pmatrix}
\gamma_B~,\\
\SS_{U_0}\gamma_A\SS_{U_0}^{-1}&=&
U_0{}_A{}^B
\gamma_B~,\\
\SS_{\hat C_0}\gamma_A \SS_{\hat C_0}^{-1}&=&\hat C_0{}_A{}^B\gamma_B~.
\eea
Since the R-R flux $F$ has  $O(1,D-1)\times O(D-1,1)$ invariance as mentioned in (\ref{ODD_map_into_solution}), the contribution of $\hat C_0$ in (\ref{hat_F_solution_beforeODOD}) vanishes.
Thus, we obtain the solution of the R-R flux $\hat F$ as
\bea
\hat F=e^{-\hat d}\SS_{\hat E^{(B)}}^{-1}\SS_{ L}\SS_{\Pi}\SS_{U_0}F^{(0)}~.
\eea

In this section, we have assumed that $F^{(0)}$ is a solution of DFT.
On the other hand, considering $F^{(0)}$ to be a solution of the DFT action 
under a fluctuation which depends on $x^m$ only, $F^{(0)}$ is, in general, not a solution 
of the action for a 'full' fluctuation depending on both $x^m$ and $\bar x_m$.
For the case that $\bar{\mathcal{G}}$ is unimodular, $F^{(0)}$ is a solution of the action for full fluctuations  
as discussed in (\ref{NSNSsector_full_fluctuation}).

As in the discussion above, a solution of the R-R flux $\check F$ in the dual space can be derived:
\bea
\sum_{p}\frac{1}{\sqrt{2^{p+1}}p!}\check F^{m_1\cdots m_p}\delta_{m_1}^{a_1}\cdots \delta_{m_p}^{a_p}\gamma_{a_1\cdots a_p}\hat K\Ket{0}&=&e^{-\check d}\SS_{\check E^{(B)}}^{-1}\SS_{\bar L}\SS_{\bar \Pi}\SS_{U_0}F^{(0)}~.
\eea
Therefore, we can derive the Poisson-Lie T-duality for the R-R sector along the line given in section \ref{sec:SUGRAvsDFT}.

\begin{enumerate}
\item	Let a solution of SUGRA be given by
\bea
\hat G+\hat B&=&L^{-1}\frac{1}{\frac{1}{G_0+B_0}+\Pi}L^{-T}~,\\
\hat\phi&=&d_0+\frac{1}{2}\log\det R+\frac{1}{4}\log\det \hat G~,\\
\hat F&=&e^{-\hat d}S_{\hat E^{(B)}}^{-1}\SS_{ L}\SS_{\Pi}\SS_{U_0}F^{(0)}~,
\eea
where $G_0,B_0,d_0$ and $F^{(0)}_{m_1\cdots m_p}$ depend only on $x^m$. Note that $\bar{\mathcal{G}}$ is unimodular.
This means that a solution of DFT is given by
\be
U=U_0~,~d=d_0~,~F=F^{(0)}~. 
\ee

\item	Transform the coordinates so that $\bar g(\bar x)g(x)=g(x')\bar g(\bar x')$.

\item	Then, we assume that the transformed solution $U_0(x(X'))$ and $F^{(0)}_{m_1\cdots m_p}(x(X'))$  depend only on $\bar x'_m$, i.e., 
\be
\frac{\partial}{\partial x'{}^m} U_0(x(X'))=0~,~\frac{\partial}{\partial x'{}^m} F^{(0)}_{m_1\cdots m_p}=0~.
\ee
and $d_0$ satisfies
\bea
d_0(x(X'))&=&d_1(x')+d_2(\bar x')~,\\
\frac{\partial}{\partial x'{}^n}(\rho(\check E_m)(d_1))&=&0~.\label{XXX}
\eea
\item Using $\check E_m=\bar R^T_m{}^aL_a{}^l\partial_l$, the condition (\ref{XXX}) means
\be
L_a{}^l\partial_ld_1=J_a~,~(J_a:\mbox{constant})~.
\ee
Define $d'_0$ and $\check I_m'$ by
\bea
d'_0(\bar x')&=&d_2(\bar x')~,\\
\check I_m'(\bar x')&=&\check I_m(\bar x')-2\rho(\check E_m)(d_1)\cr
&=&\check I_m(\bar x')-2\bar R_m^T{}^a(\bar x')J_a~,\\
\check F'&=&e^{-d_2-\frac{1}{2}\log\det \bar R}\SS_{\check E^{(B)}}^{-1}\SS_{ \bar L}\SS_{\bar \Pi}\SS_{U_0}F^{(0)}~,
\eea
Then, $d_0'$ and $\check I_m'$ do not depend on $x'$.

\item	Finally, we obtain a solution of the GSE in the dual space
\bea
\check G^{mn}+\check B^{mn}&=&\bar L^{-1}\frac{1}{G_0+B_0+\bar \Pi}\bar L^{-T}~,\\
\check \phi&=&d'_0+\frac{1}{2}\log\det \bar R+\frac{1}{4}\log\det \check G\cr
&=&d_2+\frac{1}{2}\log\det \bar R+\frac{1}{4}\log\det \check G~,\\
\check I'_m&=&\bar R^b{}_m f_{ab}{}^a-2\rho(\check E_m)(d_1)~,
\eea
\be
\sum_{p}\frac{1}{\sqrt{2^{p+1}}p!}\check F^{m_1\cdots m_p}\delta_{m_1}^{a_1}\cdots \delta_{m_p}^{a_p}
\gamma_{a_1\cdots a_p} K\Ket{0}=e^{-d_2-\frac{1}{2}\log\det \bar R}\SS_{\check E^{(B)}}^{-1}\SS_{ \bar L}\SS_{\bar \Pi}\SS_{U_0}F^{(0)}~.
\ee
Note that we use the dual vacuum $K\Ket{0}$ to define the R-R flux $\check F^{m_1\cdots m_p}$ on the dual space.
In the action of the R-R sector, $\check F^{m_1\cdots m_p}$ is contracted by the metric $\check G_{mn}$ of the dual space.

\end{enumerate}

To see how the concrete form of the R-R flux $\check F'$ in the dual space is determined, we show the dual action of the R-R sector explicitly.
First, we define the R-R flux in the dual space as in the original space by
\be
\check F=\sum_p\frac{1}{\sqrt{2^{p+1}p!}}\check F^{m_1\cdots m_p}\delta_{m_1}^{a_1}\cdots\delta_{m_p}^{a_p}\gamma_{a_1\cdots a_p}K\Ket{0}=e^{-\check d}\SS_{\check E^{(e)}}K\DGO\chi~.
\ee
Using this R-R flux $\check F$, the action of the R-R sector is given by 
\bea
S_{RR}&=&\beta_{RR}(\DGO\chi,\DGO\chi)_{AK}\cr
&=&\beta_{RR}{\big (}e^{\check d}K^{-1}\SS_{\check E^{(e)}}^{-1}\check F,e^{\check d}K^{-1}\SS_{\check E^{(e)}}^{-1}\check F{\big )}_{AK}\cr
&=&-\beta_{RR}{\big (}e^{\check d}\SS_{\check E^{(e)}}^{-1}\check F,e^{\check d}\SS_{\check E^{(e)}}^{-1}\check F{\big )}_{AK}\cr
%&=&\beta_{RR}\sum_{p,q}{\Big (}e^{\check d}K^{-1}\SS_{\check E^{(e)}}^{-1}\frac{1}{\sqrt{2^{p+1}p!}}\check F^{m_1\cdots m_p}\delta^{a_1}_{m_1}\cdots\delta^{a_p}_{m_p}\gamma_{a_1\cdots a_p}K\Ket{0},\cr
%&&~~~~~~~~~~~~~~~~~e^{\check d}K^{-1}\SS_{\check E^{(e)}}^{-1}\frac{1}{\sqrt{2^{q+1}q!}}\check F^{n_1\cdots n_q}\delta^{b_1}_{n_1}\cdots\delta^{b_q}_{n_q}\gamma_{b_1\cdots b_q}K\Ket{0}{\Big )}_{AK}\cr
&=&-\beta_{RR}\sum_{p,q}{\Big (}e^{\check d}\SS_{\check E^{(e)}}^{-1}\frac{1}{\sqrt{2^{p+1}p!}}\check F^{m_1\cdots m_p}\delta^{a_1}_{m_1}\cdots\delta^{a_p}_{m_p}\gamma_{a_1\cdots a_p}K\Ket{0},\cr
&&~~~~~~~~~~~~~~~~~e^{\check d}\SS_{\check E^{(e)}}^{-1}\frac{1}{\sqrt{2^{q+1}q!}}\check F^{n_1\cdots n_q}\delta^{b_1}_{n_1}\cdots\delta^{b_q}_{n_q}\gamma_{b_1\cdots b_q}K\Ket{0}{\Big )}_{AK}\cr
%&=&-\beta_{RR}\sum_{p,q}{\big (}e^{\check d}\SS_{\check E^{(e)}}^{-1}\frac{1}{\sqrt{2^{p+1}p!}}\check F^{m_1\cdots m_p}\delta^{a_1}_{m_1}\cdots\delta^{a_p}_{m_p}\gamma_{a_1\cdots a_p}K\Ket{0},\cr
%&&e^{\check d}\SS_{\check E^{(e)}}^{-1}\frac{1}{\sqrt{2^{q+1}q!}}\check F^{n_1\cdots n_q}\delta^{b_1}_{n_1}\cdots\delta^{b_q}_{n_q}\gamma_{b_1\cdots b_q}K\Ket{0}{\big )}_{AK}\cr
&=&-\frac{1}{2}\beta_{RR}\sum_{p}e^{2\check d}\check F^{m_1\cdots m_p}\check F^{n_1\cdots n_q}\check G_{m_1n_1}\cdots \check G_{m_pn_p}{\big (}\SS_{\check E^{(e)}}^{-1}K\Ket{0},
\SS_{\check E^{(e)}}^{-1}K\Ket{0}{\big )}_{AK}\cr
%&=&-\frac{1}{2}\beta_{RR}\sum_{p}e^{2\check d}\check F^{m_1\cdots m_p}\check F^{n_1\cdots n_q}\check G_{m_1n_1}\cdots \check G_{m_pn_p}\sqrt{\det\check G^{ll'}}(K\Ket{0},
%K\Ket{0})_{AK}\cr
&=&\frac{1}{2}\beta_{RR}\sum_{p}e^{2\check d}\check F^{m_1\cdots m_p}\check F^{n_1\cdots n_q}\check G_{m_1n_1}\cdots \check G_{m_pn_p}\sqrt{\det\check G^{ll'}}{\big (}\Ket{0},
\Ket{0}{\big )}_{AK}\cr
&=&\frac{1}{2}\beta_{RR}\sum_{p}\int dX\sqrt{\det\eta_{MN}}e^{-2d}e^{2\check d}\check F^{m_1\cdots m_p}\check F^{n_1\cdots n_q}\check G_{m_1n_1}\cdots \check G_{m_pn_p}\sqrt{\det\check G^{ll'}}\cr
&=&\frac{1}{2}\beta_{RR}\sum_{p}\int dX\det L^{-1}\sqrt{\det\check G^{ll'}}\check F^{m_1\cdots m_p}\check F^{n_1\cdots n_q}\check G_{m_1n_1}\cdots \check G_{m_pn_p}~.
\eea
On the other hand, the action of the NS-NS sector is given by (\ref{factrized_x-dependence}), in which 
the $x'$-dependence is given by a factor $\int dx'\det L^{-1}e^{-2d_1}$.
To extract this factor for the full bosonic action $S+S_{RR}$, 
we rewrite the R-R flux as 
\be
\check F'{}^{m_1\cdots m_p}=e^{d_1}\check F^{m_1\cdots m_p}~.
\ee
Using this $\check F'{}^{m_1\cdots m_p}$, the action of the R-R sector is given by
\be
S_{RR}=\frac{1}{2}\beta_{RR}\sum_{p}\int dx'\det L^{-1}e^{-2d_1}\int d\bar x'\sqrt{\det\check G^{ll'}}\check F'{}^{m_1\cdots m_p}\check F'{}^{n_1\cdots n_q}\check G_{m_1n_1}\cdots \check G_{m_pn_p}~.
\ee
Thus, the concrete form of the R-R flux $\check F'{}^{m_1\cdots m_p}$ in the dual space 
is determined as
\be
\sum_p\frac{1}{\sqrt{2^{p+1}p!}}\check F'{}^{m_1\cdots m_p}\delta_{m_1}^{a_1}\cdots\delta_{m_p}^{a_p}\gamma_{a_1\cdots a_p}K\Ket{0}=e^{d_1}e^{-\check d}\SS_{\check E^{(e)}}K\DGO\chi=e^{-d_2-\frac{1}{2}\log\det\bar R}\SS_{\check E^{(e)}}K\DGO\chi~.
\ee
To obtain the R-R flux from the algebraic structure, we consider the relation between the spin operators:
\be
\SS_{\hat E^{(B)}}\SS_{\check E^{(e)}}\SS_{\check C_0}=\SS_{\bar L}\SS_{\bar\Pi}\SS_{U_0}~,
\ee
where the spin operators $\SS_{\bar L},\SS_{\bar \Pi}$ and $\SS_{\check C_0}$ are defined by
\bea
\SS_{\bar L}\gamma_A\SS_{\bar L}^{-1}&=&
\begin{pmatrix}
\bar L{}^a{}_m&\\
&\bar L^{-T}{}_a{}^m
\end{pmatrix}
\delta_M{}^B\gamma_B~,\\
\SS_{\bar\Pi}\gamma_A\SS_{\bar\Pi}^{-1}&=&
\begin{pmatrix}
1^a{}_b&\\
\bar\Pi_{ab}&1_a{}^b
\end{pmatrix}
\gamma_B~,\\
\SS_{\check C_0}\gamma_A \SS_{\hat C_0}^{-1}&=&\hat C_0{}_A{}^B\gamma_B~,~(\check C_0\in O(1,D-1)\times O(D-1,1))~.
\eea
Using this relation, the solution of the R-R flux is given by
\bea
\sum_p\frac{1}{\sqrt{2^{p+1}p!}}\check F'{}^{m_1\cdots m_p}\delta_{m_1}^{a_1}\cdots\delta_{m_p}^{a_p}\gamma_{a_1\cdots a_p}K\Ket{0}&=&e^{-d_2-\frac{1}{2}\log\det\bar R}\SS_{\check E^{(B)}}^{-1}\SS_{\bar L}\SS_{\bar\Pi}\SS_{U_0}\SS_{\check C_0}^{-1}KF_0\cr
&=&e^{-d_2-\frac{1}{2}\log\det\bar R}\SS_{\check E^{(B)}}^{-1}\SS_{\bar L}\SS_{\bar\Pi}\SS_{U_0}\SS_{\check C_0}^{-1}F_0~,
\eea
where we used the self-duality condition: $KF_0=F_0$.
Finally, since $\check C_0\in\ O(1,D-1)\times O(D-1,1)$ is a map of a solution into another solution as discussed in the previous section, the spin operator $\SS_{\check C_0}^{-1}$ vanishes.
Thus, we obtain the solution of the R-R flux as
\bea
\sum_p\frac{1}{\sqrt{2^{p+1}p!}}\check F'{}^{m_1\cdots m_p}\delta_{m_1}^{a_1}\cdots\delta_{m_p}^{a_p}\gamma_{a_1\cdots a_p}K\Ket{0}
=e^{-d_2-\frac{1}{2}\log\det\bar R}\SS_{\check E^{(B)}}^{-1}\SS_{\bar L}\SS_{\bar\Pi}\SS_{U_0}F_0~.
\eea

\section{Discussion and outlook}

In this paper we gave a formulation of a general action for DFT obtained from a class of metric algebroid, the structure functions of which satisfy a pre-Bianchi identity, and analyzed how the gauge symmetry in $D$-dimensional theory is obtained. The action is formulated by a Lichnerowicz formula using the DGO without referring to the section condition.  
The general action contains a parameter $\beta_+$, which takes different values when we apply the formulation to the DFT${}_{sec}$ or to the DFT${}_{WZW}$ case. As a concrete example, we applied our formalism to the Poisson-Lie T-duality of the effective action on the group manifold.

We have two types of the Lichnerowicz formula, a generalized formula (\ref{generalLich}) and
a projected one  (\ref{projectLich}).\footnote{Note that the r.h.s. of (\ref{generalLich}) and  (\ref{projectLich}) can be written by the corresponding curvatures as shown in section 5. For details see \cite{2020CWMWY}.}
To obtain the generalized Lichnerowicz formula from the DGO, we need to require conditions on 
the structure functions $F_{ABC}$ and $\phi'_{ABC}$ of the metric algebroid, and also on the ambiguity of the DGO given by the flux $F_A$. These conditions are in fact the pre-Bianchi identity and a corresponding identity for $F_A$. Thus, the existence of the generalized Lichnerowicz formula is equivalent to a restriction on the metric algebroid  and a restriction on the structure of the spin bundle related to the DGO by these identities. 

The generalized Lichnerowicz formula can be interpreted as 
a sufficient condition to generate an $O(1,D-1)\times O(D-1,1)$ invariant scalar from which we obtain the projected Lichnerowicz formula given in terms of  $\mathscr L^+,\mathscr L^-$.
It is not a necessary condition, since for 
the projected Lichnerowicz formula we do not need the full pre-Bianchi identities, meaning that the parts with mixed indices $\Sa$ and $\bar\Sa$, for example $\tilde\phi_{\Sa\Sb\Sc\bar\Sd},\mathcal{B}_{\Sa\bar\Sb}$ etc., do not appear.
On the other hand, for the gauge invariance after the dimensional reduction the metric algebroid structure
is nevertheless necessary, i.e., we need the pre-Bianchi identity and therefore, we require the existence of the
generalized Lichnerowicz formula for the $2D$-dimensional space.

To recover physical models in $D$ dimensions, we use a dimensional reduction on the  fluctuations. From the point of view of physics, it is natural to require that the fluctuations should only occur in the $D$ directions which are of physical relevance. On the other hand, the geometry of the $2D$ dimensional space gives a restriction on the directions of the fluctuations. In this sense, a solution of the $2D$-dimensional background is a condition on the parameter of the $D$-dimensional field theory.

The Poisson-Lie T-duality of the $DFT_{WZW}$ has also been discussed by Hassler \cite{Hassler:2017yza}.
The coordinate transformation of the Drinfel'd Double $\mathcal{D}=\mathcal{G}\bowtie\bar{\mathcal{G}}$ and the fluctuation considered by Hassler are given by
$\Big(l(X)=\bar g(\bar x)g(x),\delta U(x)\Big)\mapsto \Big(l(X(X'))=g(x')\bar g(\bar x'),\delta U(\bar x')\Big)$ in terms of our notation.
However, this coordinate transformation does not exist in general, 
since the transformation is determined by $l(X)\mapsto l(X(X'))$. This does not necessarily mean that there exists a transformation of the fluctuation $\delta U$ satisfying the condition $\frac{\partial}{\partial x'}\delta U(x(X'))=0$. Therefore, the 'coordinate transformation' in \cite{Hassler:2017yza} is not exactly a coordinate transformation of the DFT action.

On the other hand, our procedure discussed in this paper uses
a coordinate transformation 
$\Big(l(X)=\bar g(\bar x)g(x),\delta U(X)\Big)\mapsto \Big(l(X(X'))=g(x')\bar g(\bar x'),\delta U(X')\Big)$
after extending the fluctuations to a $2D$-dimensional function, which is possible since 
we do not use a section condition. We only require the metric algebroid structure in 2D-dimensions, and that the background is a solution of DFT. Thus, the Poisson-Lie T-duality is understood as a coordinate transformation including the fluctuations.

Finally, in this paper we have also shown that starting from our general action with parametrization (\ref{action_PLTd}) we can derive the GSE of $S_{DFT_{sec}}^{mod}[E_A{}^M,d,I^m]$, which coincides the action defined in \cite{10.1093/ptep/ptx067}, thus giving the missing algebraic background for the modification of the action of the Drinfel'd Double case.

\section*{Acknowlegment}
The authors would like to thank G. Aldazabal, N. Ikeda, C. Klim\v{c}\'{i}k, Y. Sakatani, P. \v{S}evera and K. Yoshida for stimulating discussions and lectures.
We also thank T. Kaneko, S. Sekiya, S. Takezawa, and T. Yano for valuable discussions.  
S.W. is supported by the JSPS Grant-in-Aid for Scientific Research (B) No.18H01214.

\appendix
\renewcommand{\theequation}{\thesection.\arabic{equation}}

\section*{Appendix}
\section{Spin representation}
\label{spinor_representation}
Here, we construct a spin representation of $SO(10,10)$ using the basis $\Gamma_A=(\Gamma_\Sa,\Gamma_{\bar\Sa})$.
$\Gamma_A$ is defined by
\be
\Gamma_A=
\begin{pmatrix}
\Gamma_\Sa\\
\Gamma_{\bar\Sa}
\end{pmatrix}
=\frac{1}{\sqrt{2}}
\begin{pmatrix}
s_{\Sa b}&\delta_\Sa{}^{b}\\
-s_{\bar\Sa b}&\delta_{\bar\Sa}{}^b
\end{pmatrix}
\begin{pmatrix}
\gamma^b\\
\gamma_b
\end{pmatrix}~.
\ee
Note that $s=\mbox{diag}(-1,1,1,\cdots,1)$ for all suffices $a, \Sa, \bar\Sa$, 
i.e.,
\be
\Gamma^{\bar\Sa}=\eta^{\bar\Sa\bar\Sb}\Gamma_{\bar\Sb}=-s^{\bar\Sa\bar\Sb}\Gamma_{\bar\Sb}~.
\ee
The inner product of the basis is diagonalized as
\be
\{\Gamma_A,\Gamma_B\}=
2\begin{pmatrix}
s_{\Sa\Sb}&0\\
0&-s_{\bar\Sa\bar\Sb}
\end{pmatrix}~.
\ee
The component of the generalized metric using this basis is equal to that using $\gamma_A$:
\be
 H_{AB}=
\begin{pmatrix}
s_{\mathsf{ab}}&0\\
0&s_{\bar{\mathsf{a}}\bar{\mathsf{b}}}
\end{pmatrix}~.
\ee
In this paper, we use the following spin representation for the basis $\Gamma_A$:
\be
\Gamma_A=\{\Gamma_\Sa,\Gamma_{\bar\Sa}\}=\{\Gamma_{0},\cdots,\Gamma_9,\Gamma_{\bar 0},\cdots,\Gamma_{\bar 9}\}\ ,
\ee
which we give here explicitly, 
\bea
\Gamma_0&=&i\sigma_1\otimes1\otimes1\otimes1\otimes1\otimes1^{\otimes5}\cr
\Gamma_1&=&\sigma_2\otimes1\otimes1\otimes1\otimes1\otimes1^{\otimes5}\cr
\Gamma_2&=&\sigma_3\otimes\sigma_1\otimes1\otimes1\otimes1\otimes1^{\otimes5}\cr
\Gamma_3&=&\sigma_3\otimes\sigma_2\otimes1\otimes1\otimes1\otimes1^{\otimes5}\cr
\Gamma_4&=&\sigma_3\otimes\sigma_3\otimes\sigma_1\otimes1\otimes1\otimes1^{\otimes5}\cr
\Gamma_5&=&\sigma_3\otimes\sigma_3\otimes\sigma_2\otimes1\otimes1\otimes1^{\otimes5}\cr
\Gamma_6&=&\sigma_3\otimes\sigma_3\otimes\sigma_3\otimes\sigma_1\otimes1\otimes1^{\otimes5}\cr
\Gamma_7&=&\sigma_3\otimes\sigma_3\otimes\sigma_3\otimes\sigma_2\otimes1\otimes1^{\otimes5}\cr
\Gamma_8&=&\sigma_3\otimes\sigma_3\otimes\sigma_3\otimes\sigma_3\otimes\sigma_1\otimes1^{\otimes5}\cr
\Gamma_9&=&\sigma_3\otimes\sigma_3\otimes\sigma_3\otimes\sigma_3\otimes\sigma_2\otimes1^{\otimes5}\cr
\Gamma_{\bar 0}&=&(\sigma_3)^{\otimes 5}\otimes\sigma_1\otimes1\otimes1\otimes1\otimes1\cr
\Gamma_{\bar 1}&=&i(\sigma_3)^{\otimes 5}\otimes\sigma_2\otimes1\otimes1\otimes1\otimes1\cr
\Gamma_{\bar 2}&=&i(\sigma_3)^{\otimes 5}\otimes\sigma_3\otimes\sigma_1\otimes1\otimes1\otimes1\cr
\Gamma_{\bar 3}&=&i(\sigma_3)^{\otimes 5}\otimes\sigma_3\otimes\sigma_2\otimes1\otimes1\otimes1\cr
\Gamma_{\bar 4}&=&i(\sigma_3)^{\otimes 5}\otimes\sigma_3\otimes\sigma_3\otimes\sigma_1\otimes1\otimes1\cr
\Gamma_{\bar 5}&=&i(\sigma_3)^{\otimes 5}\otimes\sigma_3\otimes\sigma_3\otimes\sigma_2\otimes1\otimes1\cr
\Gamma_{\bar 6}&=&i(\sigma_3)^{\otimes 5}\otimes\sigma_3\otimes\sigma_3\otimes\sigma_3\otimes\sigma_1\otimes1\cr
\Gamma_{\bar 7}&=&i(\sigma_3)^{\otimes 5}\otimes\sigma_3\otimes\sigma_3\otimes\sigma_3\otimes\sigma_2\otimes1\cr
\Gamma_{\bar 8}&=&i(\sigma_3)^{\otimes 5}\otimes\sigma_3\otimes\sigma_3\otimes\sigma_3\otimes\sigma_3\otimes\sigma_1\cr
\Gamma_{\bar 9}&=&i(\sigma_3)^{\otimes 5}\otimes\sigma_3\otimes\sigma_3\otimes\sigma_3\otimes\sigma_3\otimes\sigma_2~,
\eea
where $\sigma_1,\sigma_2$ and $\sigma_3$ are the Pauli matrices:
\be
\sigma_1=
\begin{pmatrix}
0&1\\
1&0
\end{pmatrix}~,~
\sigma_2=
\begin{pmatrix}
0&-i\\
i&0
\end{pmatrix}~,~
\sigma_3=
\begin{pmatrix}
1&0\\
0&-1
\end{pmatrix}~.
\ee
$A_+$ is defined as a generator of the Hermitian conjugate as
\be
A_+\Gamma_AA_+^{-1}=\Gamma_A^\dagger~.
\ee
$A_+$ is the charge conjugate matrix in this representation given by
\bea
A_+=i\sigma_2\otimes\sigma_3\otimes\sigma_3\otimes\sigma_3\otimes\sigma_3\otimes \sigma_2\otimes\sigma_3\otimes\sigma_3\otimes\sigma_3\otimes\sigma_3~.
\eea
To obtain the Majorana representation, we define the generator of the 
complex conjugate as
\bea
B_+\Gamma_AB_+^{-1}&=&\Gamma_A^{*}~.
\eea
The representation of $B_+$ is given by
\bea
B_+=\sigma_3\otimes\sigma_1\otimes\sigma_2\otimes\sigma_1\otimes\sigma_2\otimes 1\otimes\sigma_2\otimes\sigma_1\otimes\sigma_2\otimes\sigma_1~.
\eea

The spinor basis $e_\alpha$ of the Majorana representation is defined by
\be
B_+^{-1}e_\alpha^*=e_{\alpha}~.
\ee
A Majorana spinor $\varphi\in\mathbb{S}$ is denoted by
\be
\varphi=\varphi^{\alpha} e_{\alpha}~.
\ee
where $\varphi^\alpha\in\mathbb{R}$.
$\Gamma _\chi$ is defined as
\be
\{\Gamma_\chi,\Gamma_A\}=0~.
\ee
The explicit form of $\Gamma_\chi$ is given by
\be
\Gamma_\chi=\sigma_3^{\otimes5}\otimes\sigma_3^{\otimes5}~.
\ee
Finally, the spin representation of the generalized metric is given by
\bea
{K}\Gamma_A{K}^{-1}&=&H_A{}^B\Gamma_B~,\\
{K}&=&1^{\otimes 5}\otimes\sigma_3^{\otimes 5}~.
\label{def_hatK}
\eea

\subsection{Majorana representation}
We list the properties under Hermitian conjugation and taking the transpose for the operators in the Majorana representation:
\bea
J_{\alpha\beta}&=&e_{\alpha}^\dagger e_{\beta}~,\\
\Gamma_A{}_{\alpha\beta}&=&e_{\alpha}^\dagger\Gamma_A e_{\beta}~,\\
A_+{}_{\alpha\beta}&=&e_{\alpha}^\dagger A_+ e_{\beta}~,\\
K_{\alpha\beta}&=&e_{\alpha}^\dagger K e_{\beta}~,
\eea
where $J_{\alpha\beta}$ is the metric on a Majorana spinor, i.e.,
\bea
e^\alpha&=&J^{-1}{}^{\alpha\beta}e_\beta~,\\
\varphi_\alpha&=&J_{\alpha\beta}\varphi^\beta~.
\eea
$J_{\alpha\beta},A_+{}_{\alpha\beta}$ and $K_{\alpha\beta}$ are (anti-)symmetric and (anti-)Hermitian, and $\Gamma_A$ is in the Majorana representation.
Then, they are real matrices as follows,
\bea
J_{\alpha\beta}^*&=&J_{\beta\alpha}~,\\
J_{\alpha\beta}&=&J_{\beta\alpha}~,\\
\Gamma_Ae_{\alpha}&=&\Gamma_{A}{}^{\beta}{}_{\alpha}e_{\beta}~,\\
\Gamma_{A}{}_{\alpha\beta}&=&\Gamma_{A}^*{}_{\alpha\beta}~,\\
A_+^*{}_{\alpha\beta}&=&-A_{+\beta\alpha}~,\\
A_+{}_{\alpha\beta}&=&- A_+{}_{\beta\alpha}~,\\
K^*{}_{\alpha\beta}&=& K_{\beta\alpha}~,\\
K{}_{\alpha\beta}&=&K{}_{\beta\alpha}~.
\eea

\subsection{Vacuum}
The vacuum of the spin space $\mathbb{S}$ is defined by
\bea
 B_+^{-1}\Ket{0}^*&=&\Ket{0}~,\cr
\gamma_a\Ket{0}&=&0~.
\label{def_S_vacuum}
\eea
$\gamma_a$ is a suffix of standard $O(D,D)$.
On the other hand, $K\Ket{0}$ is a dual vacuum and satisfies
\be
\gamma^aK\Ket{0}=0~.\label{dualvacuumofspinor}
\ee

\subsection{A-product}
An $O(D,D)$ transformation of a spinor is given by
\be
\varphi\mapsto\exp\Big(\frac{1}{4}\Lambda_{AB}\Gamma^{AB}\Big)\varphi~,
\ee
where $\Lambda_{AB}=-\Lambda_{BA}$.
We define the A-product $(\cdot,\cdot)_{A}$ as an $O(D,D)$ invariant product of any two spinors $\varphi_1,\varphi_2$:
\be
(\varphi_1,\varphi_2)_A=\varphi_1^\dagger A_+\varphi_2~.
\label{A-inner_product}
\ee
We show the $O(D,D)$ invariance of the A-product as follows,
\bea
(\varphi_1,\varphi_2)_A&\mapsto&\Big(\exp\Big(\frac{1}{4}\Lambda_{AB}\Gamma^{AB}\Big)\varphi_1,\exp\Big(\frac{1}{4}\Lambda_{A'B'}\Gamma^{A'B'}\Big)\varphi_2\Big)_A\cr
&=&\varphi_1^\dagger\exp\Big(\frac{1}{4}\Lambda_{AB}\Gamma^{AB}\Big)^\dagger  A_+\exp\Big(\frac{1}{4}\Lambda_{A'B'}\Gamma^{A'B'}\Big)\varphi_2\cr
&=&\varphi_1^\dagger A_+\exp\Big(-\frac{1}{4}\Lambda_{AB}\Gamma^{AB}\Big) \exp\Big(\frac{1}{4}\Lambda_{A'B'}\Gamma^{A'B'}\Big)\varphi_2\cr
&=&\varphi_1^\dagger A_+\varphi_2\cr
&=&(\varphi_1,\varphi_2)_A~.
\label{A-inner_product_invariance_under_ODD}
\eea

The A-product of two Majorana spinors $\varphi_1,\varphi_2$ is real:
\be
(\varphi_1,\varphi_2)_A=\varphi_1^{\alpha}e_{\alpha}^\dagger  A_+\varphi_2^{\beta}e_{\beta}=\varphi_1^{\alpha} A_{+\alpha\beta}\varphi_2^{\beta}\in \mathbb{R}~.
\ee
As we see from (Eq.4.42), $(\Ket{0},K\Ket{0})_A$ includes the measure.
The representation of $(\Ket{0},K\Ket{0})_A$ is given by
\be
(\Ket{0},K\Ket{0})_A=\Bra{0}A_+K\Ket{0}~.
\ee
Since $A_+$ and $K$ are constant matrices,
$\Ket{0}$ has to be a half density.
\be
\Ket{0}= \sqrt{dXe^{-2d}\sqrt{\det\eta_{NM}}}\Ket{\bar{0}}~,
\ee
where $\Ket{\bar{0}}$ is a constant spinor and satisfies
\be
\Bra{\bar{0}}A_+K\Ket{\bar{0}}=c_0~.
\ee

\subsection{AK-product}
We can define the AK-product $(-,-)_{AK}$ as the $O(1,D-1)\times O(D-1,1)$ invariant product of any two spinors $\varphi_1,\varphi_2$:
\be
(\varphi_1,\varphi_2)_{AK}=\varphi_1^\dagger A_+ K\varphi_2~.
\label{AK-inner_product}
\ee
We can see its $O(1,D-1)\times O(D-1,1)$ invariance as:
\bea
(\varphi_1,\varphi_2)_{AK}&\mapsto&\Big(\exp\Big(\frac{1}{4}\Lambda_{AB}\Gamma^{AB}\Big)\varphi_1,\exp\Big(\frac{1}{4}\Lambda_{A'B'}\Gamma^{A'B'}\Big)\varphi_2\Big)_{AK}\cr
&=&\varphi_1^\dagger\exp\Big(\frac{1}{4}\Lambda_{AB}\Gamma^{AB}\Big)^\dagger  A_+ K\exp\Big(\frac{1}{4}\Lambda_{A'B'}\Gamma^{A'B'}\Big)\varphi_2\cr
&=&\varphi_1^\dagger A_+\exp\Big(-\frac{1}{4}\Lambda_{AB}\Gamma^{AB}\Big)K \exp\Big(\frac{1}{4}\Lambda_{A'B'}\Gamma^{A'B'}\Big)\varphi_2\cr
&=&\varphi_1^\dagger A_+ K\exp\Big(-\frac{1}{4}\Lambda_{AB}\Gamma^{CD}H_{C}{}^AH_D{}^B\Big) \exp\Big(\frac{1}{4}\Lambda_{A'B'}\Gamma^{A'B'}\Big)\varphi_2\cr
&=&\varphi_1^\dagger A_+ K\exp\Big(-\frac{1}{4}\Lambda_{AB}\Gamma^{AB}\Big) \exp\Big(\frac{1}{4}\Lambda_{A'B'}\Gamma^{A'B'}\Big)\varphi_2\cr
&=&\varphi_1^\dagger A_+ K\varphi_2\cr
&=&(\varphi_1,\varphi_2)_{AK}~,
\label{AK-inner_product_invariance_under_ODOD}
\eea
where $\Lambda_{AB}$ satisfies
\be
\Lambda_{AB}=-\Lambda_{BA}~,~\Lambda_{AB}=H_A{}^{A'}H_B{}^{B'}\Lambda_{A'B'}~.
\ee
As an A-product, the AK-product is real.
\be
(\varphi_1,\varphi_2)_{AK}=\varphi_1^{\alpha}e_{\alpha}^\dagger  A_+ K\varphi_2^{\beta}e_{\beta}=\varphi_1^{\alpha} A_{+\alpha\gamma}S^{\gamma\delta} K_{\delta\beta}\varphi_2^{\beta}
\ee
\be
(\varphi_1,\varphi_2)_{AK}\in \mathbb{R}~.
\ee

\section{Reduction of $S_{RR}$ to DFT$_{sec}$}\label{reductiontoDFT}
In this section, we show that for the case of a flat background $S_{RR}$ reduces to DFT$_{sec}$. For this end we consider the case where the metric and the structure function $\phi'$ are given by
\be
\eta_{MN}=
\begin{pmatrix}
0&\delta^m{}_n\\
\delta_m{}^n&0
\end{pmatrix}~,~\phi'_{LM}{}^N=0~.
\ee
Then, the vielbein $E_A{}^N$ is given by
\be
E_A{}^N=
\begin{pmatrix}
e^{-T}{}^a{}_m&0\\
0&e_a{}^m
\end{pmatrix}
\begin{pmatrix}
\delta^m{}_n&0\\
-B_{mn}&\delta_m{}^n
\end{pmatrix}~.
\ee
For the purpose to define the $B$-transformation later, we separate $E_A{}^N$ into a $GL(D)$ part $E^{(e)}_A{}^M$ and a B-field part $E^{(B)}_M{}^N$ as
\bea
E_A{}^M&=&E_A^{(e)}{}^NE_N^{(B)}{}^M~,\\
E^{(e)}_A{}^N&=&
\begin{pmatrix}
e^{-T}{}^a{}_m&0\\
0&e_a{}^m
\end{pmatrix}~,\\
E^{(B)}_N{}^M&=&
\begin{pmatrix}
\delta^m{}_n&0\\
-B_{mn}&\delta_m{}^n
\end{pmatrix}~.
\eea
The spin operators $\SS_{E^{(e)}},\SS_{E^{(B)}}$ related to $E^{(e)}$ and $E^{(B)}$, respectively, can be defined as
\bea
\SS_{E^{(e)}}\gamma_A\SS_{E^{(e)}}^{-1}&=&E^{(e)}_A{}^N\delta_N{}^B\gamma_B~,\\
\SS_{E^{(B)}}\gamma_A\SS_{E^{(B)}}^{-1}&=&\delta_A{}^ME^{(B)}_M{}^N\delta_N{}^B\gamma_B~.
\eea
The concrete form of $\SS_{E^{(e)}}$ and $\SS_{E^{(B)}}$ are
\bea
\SS_{E^{(e)}}&=&\exp(-\frac{1}{2}\lambda_a{}^b\gamma^a{}_b)~,~(e^{\lambda})_a{}^b=e_a{}^m\delta_m{}^b~,\\
\SS_{E^{(B)}}&=&e^{\frac{1}{4}B_{mn}\delta_a{}^m\delta_b{}^n\gamma^{ab}}~,
\eea
Using these spin operators, we define the spin operator $\SS_E$ related to $E_A{}^N$ as
\bea
\SS_E&=&\SS_{E^{(B)}}\SS_{E^{(e)}}~,\\
\SS_E\gamma_A\SS_E^{-1}&=&E_A{}^N\delta_N{}^B\gamma_B~.
\eea
With this $\SS_E$, the corresponding Dirac generating operator $\DGO$ denoted by $\partial_N:=E_N^{-1}{}^A\partial_A$ is
\be
e^{-d}\SS_E\DGO(e^{-d}\SS_E)^{-1}=\frac{1}{2}\gamma^A\delta_A{}^N\partial_N=:\DGO_0~.
\ee
$\DGO_0$ is a Dirac operator in the local coordinate basis.
To see the relation between $S_{RR}$ and DFT$_{sec}$, 
we rewrite the action of the R-R sector $S_{RR}$ by $\DGO_0$:
\bea
S_{RR}&=&\beta_{RR}(\DGO\chi,\DGO\chi)_{AK}\cr
&=&\beta_{RR}(\DGO\chi)^\dagger  A_+ K\DGO\chi\cr
&=&\beta_{RR}((e^{-d}\SS_E)^{-1}\DGO_0(e^{-d}\SS_E)\chi)^\dagger A_+ K(e^{-d}\SS_E)^{-1}\DGO_0(e^{-d}\SS_E)\chi\cr
&=&\beta_{RR}(\SS_{E^{(B)}}^{-1}\DGO_0\breve\chi)^\dagger e^d(\SS_{E^{(e)}}^{-1})^\dagger A_+ K(e^d\SS_{E^{(e)}}^{-1}\SS_{E^{(B)}}^{-1})\DGO_0\breve\chi\cr
&=&\beta_{RR}(\SS_{E^{(B)}}^{-1}\DGO_0\breve\chi)^\dagger e^{2d} A_+\SS_{E^{(e)}} K\SS_{E^{(e)}}^{-1}(\SS_{E^{(B)}}^{-1}\DGO_0\breve\chi)~,
\eea
where $\breve \chi$ is defined by
\be
\breve\chi=e^{-d}\SS_E\chi~.
\ee
The coefficient of the R-R flux is defined by
\bea
\sum_{p}\frac{1}{\sqrt{2}^pp!}\breve{F}^{RR}_{m_1\cdots m_p}\delta^{m_1}_{a_1}\cdots \delta^{m_p}_{a_p}\gamma^{a_1\cdots a_p}\Ket{0}
&:=&\sqrt{2}\SS_{E^{(B)}}^{-1}\DGO_0\breve\chi\cr
&=&e^{-\frac{1}{4}B_{mn}\delta^m_a\delta^n_b\gamma^{ab}}\frac{1}{\sqrt{2}}\gamma^C\delta_C{}^N\partial_N\breve\chi~.
\eea
This defines the $RR$ field $\breve F^{RR}$ from the spinor $\breve \chi$. It is the equivalent to the relation used in DFT$_{sec}$ \cite{Hohm:2011aa}.
Using $\breve F^{RR}$, the action of the R-R sector becomes
\bea
S_{DFT}^{RR}&=&\beta_{RR}(\SS_{E^{(B)}}^{-1}\DGO_0\breve\chi)^\dagger e^{2d} A_+\SS_{E^{(e)}} K\SS_{E^{(e)}}^{-1}(\SS_{E^{(B)}}^{-1}\DGO_0\breve\chi)\cr
&=&\frac{1}{2}\beta_{RR}\Big(\sum_{p}\frac{1}{\sqrt{2}^pp!}\breve{F}^{RR}_{m_1\cdots m_p}\delta_{a_1}^{m_1}\cdots\delta_{a_p}^{m_p}\gamma^{a_1\cdots a_p}\Ket{0}\Big)^\dagger e^{2d} A_+\SS_{E^{(e)}} K\SS_{E^{(e)}}^{-1}\cr
&&~~~~~~~~~~~~~\Big(\sum_{q}\frac{1}{\sqrt{2}^qq!}\breve{F}^{RR}_{n_1\cdots n_q}\delta_{b_1}^{n_1}\cdots\delta_{q_1}^{n_q}\gamma^{b_1\cdots b_q}\Ket{0}\Big)\cr
&=&\frac{1}{2}\beta_{RR}\sum_{p,q}\frac{1}{\sqrt{2}^{p+q}p!q!}\breve{F}^{RR}_{m_1\cdots m_p}\breve{F}^{RR}_{n_1\cdots n_q}\delta_{a_1}^{m_1}\cdots\delta_{a_p}^{m_p}\delta_{b_1}^{n_1}\cdots\delta_{b_q}^{n_q}\Bra{0} e^{2d}\gamma^{a_1\cdots a_p}{}^\dagger A_+\SS_{E^{(e)}} K\SS_{E^{(e)}}^{-1}\gamma^{b_1\cdots b_q}\Ket{0}\cr
&=&\frac{1}{2}\beta_{RR}\sum_{p,q}\frac{1}{\sqrt{2}^{p+q}p!q!}\breve{F}^{RR}_{m_1\cdots m_p}\breve{F}^{RR}_{n_1\cdots n_q}\delta_{a_1}^{m_1}\cdots\delta_{a_p}^{m_p}\delta_{b_1}^{n_1}\cdots\delta_{b_q}^{n_q}\Bra{0} e^{2d} A_+\gamma^{a_p\cdots a_1}\SS_{E^{(e)}} K\SS_{E^{(e)}}^{-1}\gamma^{b_1\cdots b_q}\Ket{0}\cr
&=&\frac{1}{2}\beta_{RR}\sum_{p,q}\frac{1}{\sqrt{2}^{p+q}p!q!}\breve{F}^{RR}_{m_1\cdots m_p}\breve{F}^{RR}_{n_1\cdots n_q}\delta_{a_1}^{m_1}\cdots\delta_{a_p}^{m_p}\delta_{b_1}^{n_1}\cdots\delta_{b_q}^{n_q}\cr
&&\times\Bra{0} e^{2d} A_+\SS_{E^{(e)}} K\SS_{E^{(e)}}^{-1}(G^{m'_1n'_1}\delta^{a_1}_{m'_1}\delta^{c_1}_{n'_1})\cdots (G^{m'_pn'_p}\delta^{a_p}_{m'_p}\delta^{c_p}_{n'_p})\gamma_{c_p\cdots c_1}\gamma^{b_1\cdots b_q}\Ket{0}\cr
&=&\frac{1}{2}\beta_{RR}\sum_{p}\frac{1}{p!}\breve{F}^{RR}_{m_1\cdots m_p}\breve{F}^{RR}_{n_1\cdots n_p}\delta_{a_1}^{m_1}\cdots\delta_{a_p}^{m_p}\delta_{b_1}^{n_1}\cdots\delta_{b_p}^{n_p}\cr
&&\times\Bra{0} e^{2d} A_+\SS_{E^{(e)}} K\SS_{E^{(e)}}^{-1}(G^{m'_1n'_1}\delta^{a_1}_{m'_1}\delta^{b_1}_{n'_1})\cdots (G^{m'_pn'_p}\delta^{a_p}_{m'_p}\delta^{b_p}_{n'_p})\Ket{0}\cr
&=&\frac{1}{2}\beta_{RR}\sum_{p}\frac{1}{p!}\breve{F}^{RR}_{m_1\cdots m_p}\breve{F}^{RR}_{n_1\cdots n_p}G^{m_1n_1}\cdots G^{m_pn_p}\Bra{0} e^{2d} A_+\SS_{E^{(e)}} K\SS_{E^{(e)}}^{-1}\Ket{0}\cr
%&=&\frac{1}{2}\beta_{RR}\sum_{p}\frac{1}{p!}\breve{F}^{RR}_{m_1\cdots m_p}\breve{F}^{RR}_{n_1\cdots n_p}G^{m_1n_1}\cdots G^{m_pn_p}\Bra{0} e^{2d} A_+S_{E^{(e)}} KS_{E^{(e)}}^{-1}\Ket{0}\cr
&=&\frac{1}{2}\beta_{RR}\sum_{p}\frac{1}{p!}\breve{F}^{RR}_{m_1\cdots m_p}\breve{F}^{RR}_{n_1\cdots n_p}G^{m_1n_1}\cdots G^{m_pn_p}\sqrt{\det G_{ll'}}e^{2d}\Bra{0}  A_+K\Ket{0}\cr
%&=&\frac{1}{2}\beta_{RR}\sum_{p}\frac{1}{p!}\breve{F}^{RR}_{m_1\cdots m_p}\breve{F}^{RR}_{n_1\cdots n_p}G^{m_1n_1}\cdots G^{m_pn_p}\sqrt{\det G_{ll'}}e^{2d}\Bra{0}  A_+K\Ket{0}\cr
&=&\frac{1}{2}\beta_{RR}\int dX\sqrt{\det\eta_{LL'}}e^{-2d}\sum_{p}\frac{1}{p!}\breve{F}^{RR}_{m_1\cdots m_p}\breve{F}^{RR}_{n_1\cdots n_p}G^{m_1n_1}\cdots G^{m_pn_p}\sqrt{\det G_{ll'}}e^{2d}\cr
&=&\frac{1}{2}\beta_{RR}\int dX\sum_{p}\sqrt{\det G_{ll'}}\frac{1}{p!}\breve{F}^{RR}_{m_1\cdots m_p}\breve{F}^{RR}_{n_1\cdots n_p}G^{m_1n_1}\cdots G^{m_pn_p}~,
\eea
where we used
\be
\SS_{E^{(e)}}^{-1}\Ket{0}=(\det g_{mn})^\frac{1}{4}\Ket{0}~.
\ee
To compare with the action in DFT$_{sec}$,
we determine the constant $\beta_{RR}$ as 
\be
\beta_{RR}=-\frac{1}{2c_0}~,
\ee
which yields the action of R-R sector in a flat space as:
\be
S_{RR}=-\frac{1}{4}\int dX\sum_{p}\sqrt{\det G_{ll'}}\frac{1}{p!}\breve{F}^{RR}_{m_1\cdots m_p}\breve{F}^{RR}_{n_1\cdots n_p}G^{m_1n_1}\cdots G^{m_pn_p}~.
\ee
Thus, the action of the R-R sector (\ref{R-Rsector_action}) reduced to a flat background 
is consistent with the results given in the literature for DFT$_{sec}$.

\section{GSE and DFT}

The Generalized Supergravity Equations (GSE) are defined by
\bea
R^{(e)}+4\nabla^m\partial_m\phi-4|\partial\phi|^2-\frac{1}{2}|H|^2-4(I^mI_m+U^mU_m+2U^m\partial_m\phi-\nabla_mU^m)&=&0~,\cr
R^{(e)}_{mn}-\frac{1}{4}H_{mpq}H_n{}^{pq}+2\nabla_m\partial_n\phi+\nabla_mU_n+\nabla_nU_m&=&0~,\cr
-\frac{1}{2}\nabla^kH_{kmn}+\partial_k\phi H^k{}_{mn}+U^kH_{kmn}+\nabla_mI_n-\nabla_nI_m&=&0~.
\label{concreteformGSE}
\eea
Here $R^{(e)}$ is the Ricci scalar given by the $D$-dimensional vielbein and
$I=I^m\partial_m$ is a constant Killing vector which satisfies
\be
L_IG=0~,~L_IB=0~,~L_I\phi=0~.
\ee
$U_m$ is defined by
\be
U_m=I^nB_{nm}~.
\ee
It was shown in \cite{10.1093/ptep/ptx067} that the GSE can be derived from the DFT$_{sec}$ by taking an ansatz
\be
H_{MN}=
\begin{pmatrix}
G^{-1}&-G^{-1}B\\
BG^{-1}&G-BG^{-1}B
\end{pmatrix}~,~
d=\phi-\frac{1}{4}\log(\det G)+I^m\tilde x_m~.
\ee
On the other hand, we would like to use a different ansatz, 
%\be
%H_{MN}=
%\begin{pmatrix}
%G^{-1}&G^{-1}B\\
%-BG^{-1}&G-BG^{-1}B
%\end{pmatrix}~,
%\ee
to obtain the Poisson-Lie T-duality in which the sign of $B_0$ equals to that of $\Pi$ and $\bar\Pi$ as in section \ref{PLTD}
Since a change of the ansatz $(B,I)\rightarrow (-B,-I)$ respects the GSE (\ref{concreteformGSE}), we use here the ansatz 
\bea
H_{MN}=
\begin{pmatrix}
G^{-1}&G^{-1}B\\
-BG^{-1}&G-BG^{-1}B
\end{pmatrix}~,
\label{H_ansatz_GSE}
\\
d=\phi-\frac{1}{4}\log(\det G)-I^m\tilde x_m~.
\eea
Moreover, when $I^m$ is not constant, we use the following redefinition of $F_A$:
\bea
d&=&\phi-\frac{1}{4}\log(\det G)~,\cr
F_M&=&2\partial_Md-2I_M~,
\label{FM_ansatz_GSE}
\eea
where $I_M=(I^m,I_m)$ as in the modified DFT.

Thus, in this paper, we denote the DFT action ${\cal I}(0,8c_0^{-1})$ using the ans\"atze 
(\ref{H_ansatz_GSE}), (\ref{FM_ansatz_GSE}) which derives the GSE of $S_{DFT_{sec}}^{mod}[E_A{}^M,d,I^m]$. The resulting action coincides with the modified DFT action defined in \cite{Sakatani_2017}, and thus is giving the missing algebraic background of their modification for the Drinfel'd double case .

\bibliographystyle{JHEP}
\bibliography{watamura}

\end{document}